\documentclass[a4paper,12pt,reqno,twoside, titlepage]{amsart}
\usepackage[margin=2cm]{geometry}
\usepackage{setspace}
\usepackage{amsmath, mathtools, calligra}
\usepackage{amsfonts}
\usepackage{amssymb}
\usepackage{xcolor}
\usepackage{graphicx}
\usepackage{pstricks}
\usepackage{array}
\usepackage{pictexwd}
\usepackage{fancybox}
\usepackage{newcent}
\usepackage{enumitem}
\usepackage{tikz}
\usepackage{mathrsfs}
\usepackage{bm}
\usepackage{dsfont}
\usepackage{subcaption}
\setlist{nolistsep}
\newcolumntype{C}[1]{>{\centering\let\newline\\\arraybackslash\hspace{0pt}}m{#1}}
\DeclareMathAlphabet{\mathpzc}{OT1}{pzc}{m}{it}

\newtheorem{theorem}{Theorem}
\newtheorem{proposition}{Proposition}

\newtheorem{definition}{Definition}

\def \co{ {\rm co\,} }

\def \proba{\mathbb{P}}

\def \marg {\textrm{marg}}
\newcommand\indic{\mathds{1}}
\def \beq{\begin{eqnarray*}}
\def\eeq{\end{eqnarray*}}

\usepackage[textwidth=30mm]{todonotes}

\usepackage{mathtools}

\begin{document}
\title{Information Design in Multi-stage Games}
\date{\today}
\thanks{Ludovic Renou gratefully acknowledges the support of the Agence Nationale pour la Recherche under grant ANR CIGNE (ANR-15-CE38-0007-01) and through the ORA Project ``Ambiguity in Dynamic Environments'' (ANR-18-ORAR-0005). We thank Laura Doval, Stephen Morris, Sujoy Mukerji, Peter Norman, Alessandro Pavan, and Alex Wolitzky for insightful comments and the audiences at the many seminars we have given. We are particularly indebted to Tristan Tomala for his generosity with time,  pointed discussions and perspective comments. }
\author{Miltiadis Makris}
\address{Miltiadis Makris, Department of Economics, University of Kent, UK}
\email{mmakris.econ(at)gmail.com}
\author{Ludovic Renou}
\address{Ludovic Renou, Queen Mary University of London , CEPR and University of Adelaide, Miles End, E1 4NS, London, UK}
\email{lrenou.econ(at)gmail.com}

\begin{abstract}
This paper generalizes the concept of \emph{Bayes correlated equilibrium} (Bergemann and Morris, 2016) to multi-stage games. We demonstrate the power of our characterization results by applying them to a number of illustrative examples and applications.


\end{abstract}

\maketitle

\newpage

\section{Introduction}

This paper generalizes the concept of \emph{Bayes correlated equilibrium} (Bergemann and Morris, 2016) to multi-stage games. In a multi-stage game, a set of players interact over several stages and, at each stage, players receive private signals about past and current (payoff-relevant) states, past actions and past signals, and choose actions. Repeated games and, more generally, stochastic games are examples of multi-stage games.\medskip

Consider an analyst, who postulates a multi-stage game, which we call the \emph{base game}, but also acknowledges that  players may receive \emph{additional} signals, which can depend on past and current states, past actions, current and past signals (including the past additional ones). Which predictions can the analyst make if he does not want to assume a fixed \emph{expansion} of the base game, i.e., a multi-stage game that differs from the base game only in that players have fixed \emph{additional} signals? \medskip

Bergemann and Morris (2016) address that question within the class of static games. These authors show that the Bayes correlated equilibria of the (static) base game characterize all the predictions the analyst can make. (See below for an informal definition of a Bayes correlated equilibrium.) In many economic applications, however, the interaction between the economic agents is best modeled as a dynamic game, where the agents receive information over time and have the opportunity to make multiple decisions. \medskip

As an example, consider the refinancing operations of central banks. Typically, central banks organize weekly tender auctions to provide short-term liquidities to financial institutions. While extensive regulations carefully specify the auction formats central banks use, the information the financial institutions and the central banks receive over time as well as the  communication between them are substantially harder to model. An analyst may thus want to postulate a base game, which captures all that is known to him -- auction format, public annoucements, public statistics -- and to remain agnostic about the private information the financial institutions and central banks have. In other words, the analyst considers all possible \emph{expansions} of the base game. Recent contributions in the econometrics literature on partial identification have adopted such an approach. See  Bergemann, Brooks and Morris (2019), Gualdani and Sinha (2021), Magnolfi and Roncoroni (2017), Syrgkanis, Tamer, and Ziani (2018). \medskip

Our main contribution is methodological. We derive several generalizations of the concept of Bayes correlated equilibrium, where each generalization corresponds to a solution concept for multi-stage games. We focus primarily on the concept of Bayes-Nash equilibrium. While refinements are frequently used in applications, we do so for a simple reason: the logical arguments do not differ from one solution concept to another.  Our main theorem (Theorem 1) states an equivalence between (i) the set of all distributions over states and actions induced by all \emph{Bayes-Nash equilibria} of all expansions of the base game, and  (ii) the set of all distributions over states and actions induced by all \emph{Bayes correlated equilibria} of the base game.\medskip 

  At a Bayes correlated equilibrium of the base game, at each stage, an ``omniscient'' mediator makes private recommendations of actions to players, conditional on past and current states and signals, past actions and past recommendations. In other words, the mediator makes recommendations at each history of the base game.  Moreover,  at each stage, players have an incentive to be obedient, if they have never disobeyed in the past, and expect others to have been obedient in the past and to continue to be in the future.  We stress here that the ``omniscient'' mediator is a metaphor, an abstract entity, which only serves as a tool to characterize all the equilibrium outcomes we can obtain by varying the information structures. \medskip 
  
  The logical arguments are simple. Fix an expansion of the base game and an equilibrium. We show that we can emulate the equilibrium of the expansion as an equilibrium of an auxiliary mediated game, where a dummy (additional) player makes reports to a mediator and the mediator sends messages to the original players.  In that auxiliary game, the dummy player knows the actions, signals and states and the messages the mediator sends are the additional signals of the expansion. We can then apply the classical revelation principle of Myerson (1986) and Forges (1986) to replicate the  equilibrium of the mediated game as  a canonical equilibrium of the mediated game, where players are truthful and obedient, provided they have been in the past.  At that canonical equilibrium, the mediator is ``omniscient'' at truthful histories and players are obedient provided they have been in the past: we have a Bayes correlated equilibrium. The very same logic generalizes to a variety of other solution concepts. All we need is a revelation principle. \medskip

  Finally, we provide two illustrations of the broad applicability of our results.   In particular, we generalize the characterization of de Oliveira and Lamba (2019). We refer the reader to Section 5 for more details.\medskip

The closest paper to ours is Bergemann and Morris (2016), henceforth BM. These authors characterize the set of distributions over actions and states induced by all Bayes-Nash equilibria of all expansions of static base games, and show the equivalence with the distributions induced by the Bayes correlated equilibria of the static base games. The present paper generalizes their work to dynamic problems. Three insights emerge from our generalization. \medskip

The primitives in BM are a set of payoff-relevant states $\Omega$, a set of base signals $S$ and a distribution $p$ over $\Omega \times S$. There are two equivalent definitions of an expansion in static problems. The first definition states that an expansion is a set of additional signals $M$ and a joint distribution $\pi$ over $\Omega \times S \times M$ such that the marginal over $\Omega \times S$ is $p$. The second definition states that an expansion is a set of additional signals $M$ and a kernel $\xi$ from $\Omega \times S$ to probability distributions over $M$. Both definitions have natural analogues in multi-stage games, but they stop being equivalent. A first insight of our analysis is that the work of BM generalizes to dynamic problems with the latter definition, but not with the former. Intuitively,  the latter definition induces a well-defined strategy for the mediator in our auxiliary mediated game, while the former might not. (See Section 3.)  

A second insight of our analysis is that we genuinely need the mediator to make recommendations at \emph{all} histories.  To understand the need for this, note that even in dynamic games where all the states and signals about the states are drawn ex-ante, it would not be enough to have the mediator recommend strategies as a function of the realized states and signals at the first stage only. The reason is that players' signals at interim stages may also provide private information about the actions taken by players in earlier stages. For instance, if the base game is a repeated game with imperfect monitoring, a possible expansion is to perfectly inform players of past actions. As a result, if the mediator could not react to deviations that are unobserved by some players, it might not be able to induce the appropriate beliefs, and thereby actions, on the part of players in the base game. In fact, as the introductory example (Section 2) demonstrates, applying the definition of BM on the strategic form of even the simplest multi-stage games does \emph{not} characterize what we can obtain by considering all equilibria of all expansions. 

The third insight is that the analysis of BM generalizes to any solution concept for which a revelation principle holds. In particular, this is true for the two versions of perfect Bayesian equilibrium we consider. This is particularly important for many economic applications. Bargaining problems (e.g., Bergemann, Brooks and Morris, 2015), allocation problems with aftermarkets (e.g., Calzolari and Pavan, 2006, Giovannoni and Makris, 2014, and Dworczak, 2017), dynamic persuasion problems (Ely, 2017 and Renault, Solan and Vieille, 2017) are all instances of dynamic problems, where sequential rationality is a natural requirement.

\medskip

Doval and Ely (2020) is another generalization of the work of BM and nicely complements our own generalization. These authors take as given states, consequences and state-contingent payoffs over the consequences, and characterize all the distributions over states and consequences consistent with the players playing according to \emph{some} extensive-form game. Our work differs from theirs in two important dimensions. First, we take as given the base game (and, thus, the order of moves). In some economic applications, it is a reasonable assumption. For instance, if we think about the refinancing operations of central banks, the auction format and their frequencies define the base game. If a first-price auction is used to allocate liquidities, it would not make sense to consider games, where another auction format is used. In other applications, this is a more problematic assumption. For instance, if we think about Brexit and the negotiations between the European Union and the United Kingdom, it was difficult to have a well-defined base game in mind. Second, unlike Doval and Ely, we are able to accommodate dynamic problems, where the evolution of states and signals is controlled by the players through their actions. This is a natural assumption in many economic problems, such as mergers with ex-ante match-specific investments or inventory problems.\medskip

Finally, this paper contributes to the literature on correlated equilibrium and its generalizations, e.g., communication equilibrium (Myerson, 1986, Forges, 1986), extensive-form correlated equilibrium (von Stengel and Forges, 2008), or Bayesian solution (Forges, 1993, 2006).\footnote{
The concept of extensive-form correlated equilibrium was first introduced in Forges (1986). The concept introduced in von Stengel and Forges (2008) differs from the one in Forges (1986).} The concept of Bayes correlated equilibrium is a generalization of all these notions. Solan (2001) is a notable exception. In stochastic games with players perfectly informed of past actions and past and current states, Solan considers general communication devices, where the mediator sends messages to the players as a function of past messages sent and received, and the history of the game, i.e., the past actions, and the past and current states. Solan's mediator is omniscient. For that class of games, Solan shows that the set of Bayes correlated equilibrium payoffs is equal to the set of extensive-form correlated equilibrium payoffs. As we show in Example 4, this equivalence does not hold if players are not perfectly informed of past actions. See Forges (1985) for  a related result.

\section{An Introductory Example} 
This section illustrates our  main results with the help of  a simple example. The example illustrates a novel and distinctive aspect of information design in dynamic games:  In addition to providing information about payoff-relevant states, the designer can choose the information players have about the past actions of others. E.g., in voting problems, the designer can choose how much information to reveal about past votes.\medskip

\textit{\textbf{Example 1.}} There are two players, labelled 1 and 2, and two stages. Player 1 chooses either $T$ or $B$ at the first stage, while player $2$ chooses either $L$ or $R$ at the second stage. In the base game, player $2$ has no information about player $1$'s choice. Figure 1 depicts the base game, with the payoff of player 1 as the first coordinate. 

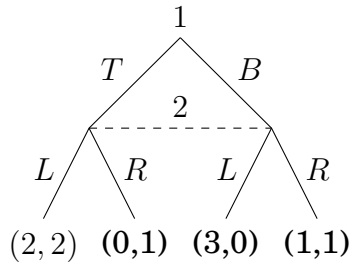
\begin{figure}[h]
\begin{center}
\begin{tikzpicture}[scale=1.2]
    \draw (1.5,2)--(0.5,1)--(0,0);
    \draw (1.5,2)--(2.5,1)--(3,0);
    \draw (2.5,1)--(2,0);
    \draw (0.5,1)--(1,0);
    \draw[dashed] (0.5,1)--(2.5,1);
  \node[below] at (0,0) {$ (2,2) $};
  \node[below] at (1,0) {(0,1)};
  \node[below] at (2,0) {(3,0)};
  \node[below] at (3,0){(1,1)};
  \node[above] at (1.5,2) {$1$};
  \node[above] at (1.5,1)  {$2$};
  \node[left] at (1,1.65) {$T$};
  \node[right] at (2,1.65) {$B$};
   \node[left] at (0.25,0.55) {$L$};
  \node[right] at (0.75,0.55) {$R$};
     \node[left] at (2.25,0.55) {$L$};
  \node[right] at (2.75,0.55) {$R$};
      \end{tikzpicture}
 \end{center}
\caption{The base game}
\end{figure}

Suppose that the information designer wants to maximize player 1's payoff.  Which information structure(s) should it design? \medskip 

To address that question, we  generalize the work of Bergemann and Morris (2016) to dynamic games. Recall that Bergemann and Morris (2016) prove that the \emph{Bayes correlated equilibria} of a static game characterize all the distributions over outcomes we can induce by varying the information structure.  At a Bayes correlated equilibrium, an omniscient mediator recommends actions, and the players have an incentive to be obedient.\medskip 

A naive idea is to apply the concept of Bayes correlated equilibrium to the strategic form of the base game. In our example, the unique Bayes correlated equilibrium of the strategic form is $(B,R)$ with a payoff profile of $(1,1)$. Working with the strategic form is, however, too restrictive. E.g., if the designer perfectly informs player 2 of player 1's action, the induced game has an equilibrium with outcome $(T,L)$ and associated payoff $(2,2)$. \medskip

Our approach is to have the omniscient mediator recommending actions to the players not only at the initial history, but at each history of the dynamic game. In addition,  the players must have an incentive to be obedient, provided they have been obedient in the past. This approach  generalizes the definition of Bayes correlated equilibria of Bergemann and Morris (2016) to multi-stage games and clearly demonstrates the need to work on the extensive-form games. (Myerson, 1986, has already pointed out the insufficiency of the strategic form; see Section 2 of his paper.) \medskip 

We now illustrate how our approach works in our example. Since the mediator is omniscient and makes recommendations at all histories, we need to consider two recommendation kernels. The first kernel  specifies the probability of recommending an action to player 1  at the first stage. The second kernel  specifies the probability of recommending an action to player 2 at the second stage as a function of the action recommended and chosen at the first stage. Players must have an incentive to be obedient. We claim that there exist such recommendation kernels with a payoff profile of $(5/2,1)$. \medskip 

To see this, assume that the mediator recommends with probability 1/2 player $1$ to play $T$ and with the complementary probability to play $B$ at the first stage, and recommends player $2$ to play $L$ at the second stage if and only if player $1$ was obedient at the first stage.  We now prove that the players have an incentive to be obedient. \medskip

If player $2$'s recommendation is $L$, he believes that player $1$ has played $T$ with probability $1/2$ and thus expects a payoff of $1$ if he plays $L$. He therefore has an incentive to be obedient. If player $2$'s recommendation is $R$, we are off the equilibrium path and any conjecture that puts probability of at least $1/2$ on player $1$ having played $B$ makes $R$ optimal. As for player $1$, he clearly has an incentive to be obedient when his recommendation is $B$ since he gets his highest payoff. When his recommendation is $T$, a deviation to $B$ is unprofitable because this leads player $2$ to play $R$. Thus, we indeed have a Bayes correlated equilibrium with a payoff profile of $(5/2,1)$. \medskip 

To answer our initial question, we now argue that no information structures can give player 1 a payoff higher than $5/2$. Since player $2$ can always play $R$, player $2$'s payoff cannot be lower than $1$. Therefore,  within the set of feasible payoff profiles, conditional on player $2$'s getting a payoff of at least $1$, player $1$'s highest payoff is $5/2$. See Section \ref{sec:4.1} for a complete characterization of the Bayes correlated equilibria of our example.\medskip

Finally, we now explain how we can use the Bayes correlated equilibrium to design an information structure, whose associated expansion generates an equilibrium payoff of $(5/2,1)$. The idea is simple: Think of recommendations as signals. Accordingly, suppose that there are two equally likely signals, $t$ and $b$, at the first stage, and two signals $l$ and $r$ at the second stage. Player 1 privately observes the first signal, while player 2 observes the signal $l$ if only if either $(t,T)$ or $(b,B)$ is the profile of signal and action at the first stage. With such information structure, players have an incentive to play according to their signals and, thus, we obtain the payoff profile $(5/2,1)$.\medskip

\section{Multi-stage Games and Expansions} 

The model follows closely Myerson (1986). There is a set $I$ of $n$ players, who interact over $T<+\infty$  stages, numbered $1$ to $T$. (With a slight abuse of notation, we denote $T$ the set of stages.) At each stage, a payoff-relevant state is drawn, players receive private signals about past and current states, past private signals and actions, and choose an action. We are interested in characterizing the joint distributions over profiles of states, actions and signals, which arise as equilibria of ``expansions'' of the game, i.e., games where players receive additional signals.  

\subsection{The base game} 

We first define the base game $\Gamma$, which corresponds to the game being played if no additional signals are given to the players.  At each stage $t$, a state $\omega_t \in \Omega_t$ is drawn, player $i \in I$ receives the private signal $s_{i,t} \in S_{i,t}$, which may depend probabilistically on the current and past states, past signals and actions, and then chooses an action $a_{i,t} \in A_{i,t}$. All sets are non-empty and finite.\medskip

We now introduce some notations. We write $A_t:=\times_{i \in I}A_{i,t}$ for the set of actions at stage $t$ and $A:=\times_{t\in T}\times_{i\in I}A_{i,t}$ for the set of profiles of actions. We let $H_{i,t}=A_{i,t-1} \times S_{i,t}$ be the set of player $i$'s new information at the beginning of stage $t \in \{2,\dots,T\}$, $H_{i,1} = S_{i,1}$ the set of initial information, and $H_{i,T+1}=A_{i,T}$ the set of terminal information.\medskip 

We denote $p_1(h_1,\omega_1)$ the joint probability of $(h_1,\omega_1)$ at the beginning of the first stage and $p_{t+1}(h_{t+1},\omega_{t+1}|a_t,h^t,\omega^t)$ the joint probability of  $(h_{t+1},\omega_{t+1})$ at stage $t+1$ given that $a_t$ is the profile of actions played at stage $t$ and $(h^t,\omega^t)$ is the history of actions played, signals received and states realized at the beginning of stage $t$. We assume perfect recall and, therefore, impose that $p_{t+1}((b_t,s_{t+1}),\omega_{t+1}|a_t,h^t,\omega^t)= 0$ if $b_t \neq a_t$.\medskip

We denote $H\Omega$ the subset of $\times_{t=1}^{T+1}(\times_{i \in I}H_{i,t} \times \Omega_t)$ that consists of all terminal histories of the game, with generic element $(h,\omega)$.\footnote{The sets $S_{T+1}$ and $\Omega_{T+1}$ are defined to be a singleton.}  The history $(h,\omega)$ is  in  $H\Omega$ if and only if there exists a profile of actions $a \in A$ such that 
\[p^a(h,\omega):=p_1(h_1,\omega_1) \cdot \prod_{t \in T} p_{t+1}(h_{t+1},\omega_{t+1}|a_t,h^t,\omega^t)>0.\]

For any vector $(h,\omega)$, we can denote various sub-vectors:  $h_{i}= (h_{i,1},\dots,h_{i,t},\dots, h_{i,T+1})$ the private (terminal) history of player $i$, $h_{i}^t= (h_{i,1},\dots,h_{i,t})$ the private history of player $i$ at stage $t$, $h_t=(h_{1,t},\dots,h_{n,t})$ the profile of actions played at stage $t-1$ and signals received at stage $t$, $h^t=(h_1,\dots,h_t)$ the history of signals and actions at stage $t$, $\omega= (\omega_1,\dots,\omega_T)$ the profile of realized states, and $\omega^t=(\omega_1,\dots,\omega_t)$ the profile of  states realized up to stage $t$, with corresponding sets $H_i=\{h_i: (h,\omega) \in H\Omega \mbox{\;for some\;} \omega\}$, $H_i^t=\{h_i^t: (h,\omega) \in H\Omega \mbox{\;for some\;} \omega\}$, $H_t=\{h_t: (h,\omega) \in H\Omega \mbox{\;for some\;} \omega\}$, $H^t=\{h^t: (h,\omega) \in H\Omega \mbox{\;for some\;} \omega\}$, $\Omega=\{\omega: (h,\omega) \in H\Omega \mbox{\;for some\;} h\}$,  $\Omega^t=\{\omega^t: (h,\omega) \in H\Omega \mbox{\;for some\;} h \}$. We write $H^t\Omega^t$ for the restriction of $H\Omega$ to the first $t$ stages. We let $\widehat{H}:= \times_{i \in I} H_i$ and $\widehat{H}^t:= \times_{i \in I} H^t_i$. Similar notations will apply to other sets. If there is no risk of confusion, we will not formally define these additional notations.\medskip

The payoff to player $i$ is $u_i(h,\omega)$ when the terminal history is $(h,\omega) \in H\Omega$.  We assume that payoffs do not depend on the signal realizations, i.e., for any two histories $h=(a,s)$ and $h'=(a',s')$ such that $a=a'$, $u_i(h,\omega)=u_i(h',\omega)$ for all $\omega$, for all $i$.\footnote{This is without loss of generality as we can always redefine the states to include the signals.} Throughout, we refer to the signals in $S$ as the \emph{base} signals. 

\subsection{Expansions}

In an \emph{expansion} of the base game, at each stage, players receive additional signals, which may depend probabilistically on past and current states, past and current signals (including the past additional ones), and past actions. Thus, players can receive additional information not only about the realization of current and past (payoff-relevant) states (such as the valuations for objects in auction problems) but also about the past realization of actions (as in repeated games with imperfect monitoring). Throughout, we use the same notation as in the base game to denote relevant sub-vectors and their corresponding sets. \medskip

Formally, an expansion is a collection of sets of additional private signals $(M_{i,t})_{i,t}$ and probability kernels $(\xi_t)_t$ such that all sets of additional signals are non-empty and finite, $\xi_1: H_1 \times \Omega_1 \rightarrow \Delta(M_1)$, and $\xi_t: H^t \times M^{t-1} \times \Omega^t \rightarrow \Delta(M_t)$ for all $t \geq 2$.\footnote{The set  $M_{i,T+1}$ is a singleton.} Intuitively, at each stage $t$, player $i$ receives the additional private signal $m_{i,t} \in M_{i,t}$, with \[
 \xi_{t}(m_{t}|h^{t},m^{t-1},\omega^{t})
 \]
the probability of $m_{t}$ when $(h^{t},m^{t-1},\omega^{t})$ is the history of actions, base signals, states, and past additional signals at the beginning of stage $t$. We write $M$ for the collection $(M_{i,t})_{i,t}$ and $\xi$ for $(\xi_t)_{t}$.

\medskip
Together with the base game $\Gamma$, an expansion $(M,\xi)$ induces a multi-stage, where at each stage $t$, a payoff-relevant state $\omega_t$ is realized, player $i$ receives the private signal $(s_{i,t},m_{i,t})$ and takes an action $a_{i,t}$. To complete the description of the induced multi-stage game, we let $\pi_1(h_1,m_1,\omega_1):=\xi_1(m_1|h_1,\omega_1)p_{1}(h_1,\omega_1)$ be the probability of  $(h_1,m_1,\omega_1)$ at the first stage and \[\pi_{t+1}(h_{t+1},m_{t+1},\omega_{t+1}|a_t,h^t,m^t,\omega^t):=\xi_{t+1}(m_{t+1}|h^{t+1},m^t,\omega^{t+1})p_{t+1}(h_{t+1},\omega_{t+1}|a_{t},h^t,\omega^t)\] the probability of $(h_{t+1},m_{t+1},\omega_{t+1})$, when $a_t$ is the profile of actions played at stage $t$ and $(h^t,m^t,\omega^t)$ is the history of actions, signals and states at the beginning of stage $t$. With a slight abuse of language, we use the word ``expansion'' to refer to the collection of additional signals and kernels $(M, \xi)$ as well as to the multi-stage game $\Gamma_{\pi}$ induced by it.\medskip

We denote $HM\Omega$ the set of all terminal histories with $(h,m,\omega) \in HM\Omega$ if and only if there exists a profile of actions $a \in A$ such that 
 \[\pi^a(h,m,\omega):= \pi_1(h_1,m_1,\omega_1) \cdot \prod_{t \in T} \pi_{t+1}(h_{t+1},m_{t+1},\omega_{t+1}|a_t,h^t,m^t,\omega^t) >0. \] We stress that the set of terminal histories $HM\Omega$ depends on the expansion chosen. Thus, for a fixed base game and fixed sets of additional messages, different kernels $(\xi_t)_t$ may induce different sets of terminal histories $HM\Omega$. In particular, if $\xi_{t+1}(m_{t+1}|h^{t+1},m^t,\omega^{t+1})=0$, then the history $(h^{t+1},m^{t+1},\omega^{t+1})$ of  signals, actions and states up to period $t+1$ is not part of any terminal history in  $HM\Omega$.\medskip

 In closing, it is worth noting that an expansion $\xi$ induces a collection of  kernels $\pi$, with the property that $\text{marg}_{H\Omega}\pi^a = p^a$ for all $a \in A$, that is,
\begin{align*}
\sum_{(m_1,\dots,m_T)} \Big(\pi_1(h_1,m_1,\omega_1) \cdot \prod_{t \in T} \pi_{t+1}(h_{t+1},m_{t+1},\omega_{t+1}|a_t,h^t,m^t,\omega^t)\Big)  = \\
p_1(h_1,\omega_1) \cdot \prod_{t \in T} p_{t+1}(h_{t+1},\omega_{t+1}|a_t,h^t,\omega^t). \tag{$\dagger$}
\end{align*}
We call this property consistency. In static problems, the converse is also true, i.e., any consistent kernel $\pi$ induces an expansion $\xi$. However, this equivalence breaks down in dynamic problems, as the following example illustrates. \medskip

\textit{\textbf{Example 2.}} There are two stages, $\Omega_1=\Omega_2=\{0,1\}$, 
and no private signals and actions, i.e., $A_1$, $S_1$, $A_2$ and $S_2$  are singletons (for simplicity, we omit them).  The states are uniformly and independently distributed, that is,  $p_1(\omega_1) = p_2(\omega_2|\omega_1)=1/2$ for all $(\omega_1,\omega_2)$. Consider now the following sets of additional signals and kernels: $M_1=\{0,1\}$, $M_2$ is a singleton, $\pi_1(m_1,\omega_1)=1/4$  for all $(m_1,\omega_1)$ and $\pi_2(\omega_2|m_1,\omega_1)=1$ if and only if $\omega_2=(\omega_1 + m_1) \pmod 2$.  We can think of the second-stage state as the first-stage state plus a shock. 

We now verify consistency. We have
\beq 
\sum_{m_1}\pi_1(m_1,\omega_1)\pi_2(\omega_2|m_1,\omega_1) = \\
\pi_1((\omega_2- \omega_1)\pmod 2,\omega_1)\pi_2(\omega_2|(\omega_2- \omega_1)\pmod 2,\omega_1) = 1/4.
\eeq
The collection of kernels is therefore consistent. Yet, there are no kernels $(\xi_1,\xi_2)$ such that $\pi_1(m_1,\omega_1) = \xi_1(m_1|\omega_1)p_1(\omega_1)$ and $\pi_2(\omega_2|m_1,\omega_1)=p_2(\omega_2|\omega_1)$ for all $(m_1,m_2,\omega_1,\omega_2)$. 
The issue here is that $m_1$ is not just an additional piece of information about the first-stage state $\omega_1$;  it actually causes the second-stage state.  \medskip

It is worth noting that the precise definition of the base game  is important in determining whether a consistent kernel induces an expansion or not. To demonstrate this, suppose that both states are drawn at the first stage, so that the additional signal $m_1$ can now be made contingent on the new first-stage state $\omega^*_1=(\omega_1,\omega_2)$. With this reinterpretation of the base game, we have a well-defined expansion, and our analysis applies. For completeness, we formally write down this alternative multi-stage representation. Let $\Omega^*_1:= \Omega_1 \times \Omega_2$; all other sets except $M_1$ are singletons. Let $p^*_1(\omega_1,\omega_2)=1/4$ for all $(\omega_1,\omega_2)$ and $\pi^*_1(m_1,(\omega_1,\omega_2))=1/4$ if and only if $\omega_2=(\omega_1 + m_1) \pmod 2$. Finally, observe that  if we let $\xi^*_1(m_1|(\omega_1,\omega_2))= 1$ if and only if   $m_1=(\omega_2 - \omega_1) \pmod 2$, then $\pi^*_1(m_1,(\omega_1,\omega_2))= \xi_1(m_1|(\omega_1,\omega_2))p^*_1(\omega_1,\omega_2)$ for all $(m_1,\omega_1,\omega_2)$. However, such a reinterpretation is not always possible (see, for instance, Example 3).
\hfill $\blacksquare$

\section{Equivalence theorems}
This section contains our main results. It provides characterization theorems, which differ by the solution concepts adopted. In section 4.1, we  consider first the concept of Bayes-Nash equilibrium. This allows us to present our first characterization theorem in the simplest possible terms, without cluttering the analysis with issues such as consistency of beliefs, sequential rationality, or truthfulness and obedience at off-equilibrium path histories. As we will see, the main arguments extend almost verbatim to other solution concepts. In addition, if we are interested in proving an impossibility result, e.g., whether efficiency obtains, the weaker the solution concept, the stronger the result.  In section 4.2, we then extend our analysis to two refinements of the concept of Bayes-Nash equilibrium, which all impose sequential rationality.  
\subsection{A first equivalence theorem}\label{sec:4.1}

We first define the concepts of Bayes-Nash equilibrium  and Bayes correlated equilibrium. Throughout, we fix an expansion $\Gamma_{\pi}$ of $\Gamma$. \medskip

\textit{\textbf{Bayes-Nash equilibrium.}} A behavioral strategy $\sigma_i$ is a collection of maps $(\sigma_{i,t})_{t \in T}$, with $\sigma_{i,t}: H_i^tM_{i}^t \rightarrow \Delta(A_{i,t})$.  A profile $\sigma$ of behavioral strategies is a  \emph{Bayes-Nash equilibrium} of $\Gamma_{\pi}$ if

\[\sum_{\bm{h},\bm{m},\bm{\omega}}u_i(\bm{h},\bm{\omega})\proba_{\sigma,\pi}(\bm{h},\bm{m},\bm{\omega}) \geq \sum_{\bm{h},\bm{m},\bm{\omega}}u_i(\bm{h},\bm{\omega})\proba_{(\sigma'_i,\sigma_{-i}),\pi}(\bm{h},\bm{m},\bm{\omega}), \]
for all $\sigma_i'$, for all $i$, with $\proba_{\tilde{\sigma},\pi}\in \Delta(HM\Omega)$ denoting the distribution over profiles of actions, signals and states induced by $\tilde{\sigma}$ and $\pi$. We let $\mathcal{BNE}(\Gamma_{\pi})$ be the set of distributions over $H \Omega$ induced by the Bayes-Nash equilibria of $\Gamma_{\pi}$.  \medskip 

 We now state formally the main objective of our paper: we want to provide a characterization of the set $ \bigcup_{\Gamma_{\pi} \text{\,an  expansion of\;\,} \Gamma}\mathcal{BNE}(\Gamma_{\pi})$,  i.e., we want to characterize the distributions over the outcomes $H\Omega$ of the base game $\Gamma$ that we can induce by means of some expansion $\Gamma_{\pi}$ of the base game, \emph{without} any reference to particular expansions. To do so, we need to introduce the concept of Bayes correlated equilibrium of $\Gamma$. 

\medskip

\textit{\textbf{Bayes correlated equilibrium.}} Consider the following mediated extension of $\Gamma$, denoted $\mathcal{M}(\Gamma)$. At each period $t$, 
player $i$ observes the private signal $h_{i,t}$, receives a private recommendation $\hat{a}_{i,t}$ from an ``omniscient'' mediator and chooses an action $a_{i,t}$. 
We let $\tau_{i,t}: H_{i}^t \times A_{i}^{t}  \rightarrow \Delta(A_{i,t})$ be an action strategy at period $t$ and  write
$\tau_{i,t}^*$ for the obedient strategy. It is best to view the ``omniscient'' mediator as an abstraction, which makes it possible to characterize all distributions over outcomes that an information designer can induce.

A Bayes correlated equilibrium is a collection of recommendation kernels $\mu_t: H^t\Omega^t \times A^{t-1} \rightarrow \Delta(A_t)$ such that 
$\tau^*$ is an equilibrium of the mediated game, that is, 
\beq 
\sum_{\bm{h},\bm{\omega, \bm{\hat{a}}}}u_i(\bm{h},\bm{\omega})\proba_{\mu \circ \tau^*,p}(\bm{h},\bm{\omega}, \bm{\hat{a}})
\geq \sum_{\bm{h},\bm{\omega},\bm{\hat{a}}}u_i(\bm{h},\bm{\omega})\proba_{\mu \circ (\tau_i,\tau^*_{-i}),p}(\bm{h},\bm{\omega},\bm{\hat{a}})
\eeq
for all $\tau_i$, for all $i$, with $\proba_{\mu \circ \tilde{\tau},p}$ denoting the distribution over profiles of actions, base signals, states and recommendations induced by $\mu \circ \tilde{\tau}$ and $p$.  We let $\mathcal{BCE}(\Gamma)$ be the set of distributions over $H\Omega$ induced by the Bayes correlated equilibria of $\Gamma$. The set  $\mathcal{BCE}(\Gamma)$ is convex. \medskip

It is instructive to compare the concept of Bayes correlated equilibrium and communication equilibrium (Forges, 1986, Myerson, 1986). In a communication equilibrium, the mediator relies on the information provided by the players to make recommendations, while in a Bayes correlated equilibrium it is \emph{as if} the mediator  knows the realized states, actions and base signals prior to making recommendations. Let $\mathcal{CE}(\Gamma)$ be the distributions over $H\Omega$ induced by the communication equilibria of $\Gamma$. For all multi-stage games $\Gamma$, we have that $\mathcal{CE}(\Gamma) \subseteq \mathcal{BCE}(\Gamma)$ since the omniscient mediator can always replicate the Forges-Myerson mediator. Since we also have that $\mathcal{BNE}(\Gamma) \subseteq \mathcal{CE}(\Gamma)$,  we have the inclusion $\mathcal{BNE}(\Gamma) \subseteq \mathcal{BCE}(\Gamma)$. However, it is a priori not clear whether $\mathcal{BNE}(\Gamma_{\pi}) \subseteq  \mathcal{BCE}(\Gamma)$ for all expansions $\Gamma_{\pi}$ of $\Gamma$ since players have additional signals in $\Gamma_{\pi}$, while the omniscient mediator of $\Gamma$ has no additional signals. A consequence of our main result, Theorem \ref{th:equiv}, is that it is indeed the case.

\begin{theorem}\label{th:equiv} We have the following equivalence:
\[\mathcal{BCE}(\Gamma)=\bigcup_{\Gamma_{\pi} \text{\,an  expansion of\;\,} \Gamma}\mathcal{BNE}(\Gamma_{\pi}).\]
\end{theorem}
\medskip

Theorem  \ref{th:equiv} states an equivalence between (i) the set of distributions over actions, base signals and states induced by all Bayes correlated equilibria of $\Gamma$ and (ii) the set of distributions over actions, base signals and states we can obtain by considering \emph{all} Bayes-Nash equilibria of \emph{all}  expansions of $\Gamma$.\footnote{In a supplementary document, we also prove the equivalence with the set of distributions over actions, base signals and states we can obtain by considering \emph{all} communication equilibria of \emph{all}  expansions of $\Gamma$.} It is  a \emph{revelation principle} for information design. Indeed, Theorem \ref{th:equiv} states that any distribution over actions, base signals and states a designer can implement by committing to an  information structure is a Bayes correlated equilibrium distribution of the base game. We can therefore focus on the Bayes correlated equilibrium distributions and abstract from the particular information structures implementing them. This  mirrors the focus on incentive compatible social choice functions in mechanism design theory. \medskip

Theorem \ref{th:equiv} generalizes the work of BM to multi-stage games.  It is also worth emphasizing again that our definition of a Bayes correlated equilibrium is weaker than applying the definition of BM on the strategic form of the base game, which would amount to making recommendations of strategies at the first stage, as a function of the realized states and base signals. See the introductory example for an illustration. \medskip

Unlike BM's constructive proof, our proof is non-constructive.\footnote{We provide a fully constructive proof in the supplementary material.} This approach has two main advantages: (i) it reveals the main logical arguments, which are somewhat hidden in constructive proofs, and (ii) its generalization to many other solution concepts is straightforward. The central arguments are the following. Consider an expansion $\Gamma_{\pi}$ of $\Gamma$ and and equilibrium distribution  $\boldsymbol{\mu}^d \in \mathcal{BNE}(\Gamma_{\pi})$. By definition, there exists a Bayes-Nash equilibrium $\sigma$ of $\Gamma_{\pi}$, which induces $\boldsymbol{\mu}^d$.  The main idea is to replicate the expansion $\Gamma_{\pi}$ and its equilibrium $\sigma$ as a Bayes-Nash equilibrium of an auxiliary  mediated game $\mathcal{M}^*(\Gamma)$, which we now describe. The game $\mathcal{M}^*(\Gamma)$ has one additional player, called player 0, and a Forges-Myerson mediator. Player 0 is a dummy player. At the first stage,  Nature draws $(h_1,\omega_1)$ with probability $p_1(h_1,\omega_1)$, player $i$ observes $h_{i,1}$  and player $0$ observes $(h_1,\omega_1)$. Player $0$ then reports $(\hat{h}_1,\hat{\omega}_1)$ to the mediator; all other players do not make reports. The mediator then draws the message $m_1$ with probability $\xi_1(m_1|\hat{h}_1,\hat{\omega}_1)$ and sends $m_{i,1}$ to player $i$. Player 0 does not receive a message. Finally, player $i$ chooses an action $a_{i,1}$; player 0 does not take an action.  Consider now a history $(a_{t-1},h^{t-1},\omega^{t-1})$ of past actions, signals and states and a history $((\hat{h}^{t-1},\hat{\omega}^{t-1}),m^{t-1})$ of reports and messages. Stage $t$ unfolds as follows:

\begin{itemize}
\item[-] Nature draws $(h_t,\omega_t)$ with probability $p_t(h_t,\omega_t|a_{t-1},h^{t-1},\omega^{t-1})$.
\item[-] Player $i \in I $ observes the signal $h_{i,t}$ and player $0$ observes $(h_t,\omega_t)$. 
\item[-] Player $0$ reports $(\hat{h}_t,\hat{\omega}_t)$ to the mediator. All other players do not make reports. 
\item[-] The mediator draws the message $m_t$ with probability $\xi_t(m_t|\hat{h}^t,m^{t-1}, \hat{\omega}^t)$ and sends the message $m_{i,t}$  to player $i$. Player 0 does not receive a message.
\item[-] Player $i$ takes an action  $a_{i,t}$. Player 0 does not take an action.
\end{itemize}

If player 0 is truthful and each player $i \in I $  follows $\sigma_i$, we clearly have a Bayes-Nash equilibrium of the mediated game $\mathcal{M}^*(\Gamma)$, with equilibrium distribution $\boldsymbol{\mu}^d$. From the revelation principle of Forges (1986) and Myerson (1986), there exists a canonical equilibrium, where the mediator recommends actions and players are truthful and obedient provided that they have been in the past, which implements $\boldsymbol{\mu}^d$. At truthful histories, the mediator is omniscient and players have an incentive to be obedient provided they have been in the past: this is the Bayes correlated equilibrium.\medskip 

 Before applying Theorem \ref{th:equiv}, two additional remarks are worth making.  First, the above arguments are not limited to the concept of Bayes-Nash equilibrium. The same arguments apply to all solution concepts, such as weak perfect Bayesian equilibrium or conditional probability perfect Bayesian equilibrium, which admit a revelation principle. We formally state these equivalences below. Second, the above arguments clearly demonstrate the role our definition of an expansion plays. It makes it possible for the mediator to replicate any expansion as the kernels $\xi_t$ are assumed measurable with respect to the mediator's histories. With the alternative and weaker definition of an expansion as a consistent information structure, i.e., $\marg\,  \pi^a=p^a$ for all $a$, it is no longer guaranteed that the mediator, despite being omniscient, can simulate any expansion, as the next example illustrates. If at all possible, we would need an even more powerful mediator. \medskip

\textit{\textbf{Example 3.}} This example is an elaboration on Example 2. The main difference is  that the second-stage state $\omega_2$ is partially controlled by a single player through his first-stage action $a_1$.

We first define the base game. There are a single player,  two stages, two actions $A_1=\{0,1\}$ at the first stage, two states $\Omega_2=\{0,1\}$ at the second stage, and all other sets are singletons. The probabilities are: $p_2(\omega_2=1|a_1=1) =5/6$ and $  p_2(\omega_2=1|a_1=0)=1/2$. The player's payoff is one (resp., zero) if the second-stage state is zero (resp., one), regardless of his action.\smallskip

Consider now the following  information structure: $M_1=\{0,1\}$, $M_2$ is a singleton, $\pi_1(m_1=1)=1/2$, $\pi_2(\omega_2=1|a_1=1,m_1=1) =2/3$, $\pi_2(\omega_2=1|a_1=0,m_1=1) =1$, $\pi_2(\omega_2=1|a_1=1,m_1=0) =1$, and $\pi_2(\omega_2=1|a_1=0,m_1=0) =0$. This information structure is consistent, but as in Example 2, there are no kernels $(\xi_1,\xi_2)$ that induce this information structure from the base game.\smallskip
 
Player's optimal payoff is $2/3$ in the game $\Gamma_{\pi}$: the optimal strategy consists in playing $a_1=1$ (resp., $a_1=0$) when $m_1=1$ (resp., $m_1=0$). The player's optimal strategy consists in choosing the action that maximizes the likelihood of the second-stage state being $0$. The induced distribution $\mu$ over actions and states is $\mu(a_1=0,\omega_2=0)=1/2$, $\mu(a_1=0,\omega_2=1)=0$, 
$\mu(a_1=1,\omega_2=0)=1/6$, $\mu(a_1=1,\omega_2=1) =1/3$. This is not a Bayes correlated distribution. In any Bayes correlated equilibrium, the probability of $(a_1,\omega_2)$ is $\overline{\mu}_1(a_1)p_2(\omega_2|a_1)$ and there is no $\overline{\mu}_1$ that induces the distribution $\mu$. \hfill $\blacksquare$
\medskip

We close this section with an example illustrating how we can apply Theorem \ref{th:equiv}. \medskip

\textit{\textbf{Example 4.} } There are two players and two stages. Player 1 is active in the first stage and chooses an action $a_1 \in A_1$; player $2$ is inactive. Player 2 is active in the second stage and chooses an action $a_2 \in A_2$; player $1$ is inactive. There are no base signals and states, i.e., $\Omega_1$, $S_1$, $\Omega_2$ and $S_2$ are singletons. We are interested in characterizing the distributions $\mu \in \Delta(A_1 \times A_2)$ as we vary the information players have. In particular, this implies varying the information player $2$ has about the action chosen by player $1$ before choosing his own action. Formally, we consider expansions $(\xi_1,\xi_2)$, where $\xi_1 \in \Delta(M_1)$ and $\xi_2: M_1 \times A_1 \rightarrow \Delta(M_2)$. In words, player $1$ receives the additional signal $m_1$ at the first period and player 2 receives the additional signal $m_2$ at the second stage, which may depend on the first-period signal and action $(m_1,a_1)$.  

From Theorem \ref{th:equiv}, we can restrict attention to the Bayes correlated equilibria of the game. By definition, $(\mu_1,\mu_2)$ is a Bayes correlated equilibrium if:  
\beq 
 \sum_{a_1,a_2,\hat{a}_1,\hat{a}_2}u_1(a_1,a_2)\mu_1(\hat{a}_1)\tau_1^*(a_1|\hat{a}_1)\mu_2(\hat{a}_2|a_1,\hat{a}_1)\tau^*_2(a_2|\hat{a}_2) \geq \\
 \sum_{a_1,a_2,\hat{a}_1,\hat{a}_2}u_1(a_1,a_2)\mu_1(\hat{a}_1)\tau_1(a_1|\hat{a}_1)\mu_2(\hat{a}_2|a_1,\hat{a}_1)\tau^*_2(a_2|\hat{a}_2), 
\eeq
for all $\tau_1: A_1 \rightarrow \Delta(A_1)$, and 
\beq
 \sum_{a_1,a_2,\hat{a}_1,\hat{a}_2}u_2(a_1,a_2)\mu_1(\hat{a}_1)\tau_1^*(a_1|\hat{a}_1)\mu_2(\hat{a}_2|a_1,\hat{a}_1)\tau^*_2(a_2|\hat{a}_2) \geq \\
 \sum_{a_1,a_2,\hat{a}_1,\hat{a}_2}u_2(a_1,a_2)\mu_1(\hat{a}_1)\tau^*_1(a_1|\hat{a}_1)\mu_2(\hat{a}_2|a_1,\hat{a}_1)\tau_2(a_2|\hat{a}_2), 
\eeq
for all $\tau_2:A_2 \rightarrow \Delta(A_2)$, with  $\tau^*_1$ and $\tau^*_2$ the obedient strategies. Any Bayes correlated equilibrium 
$(\mu_1,\mu_2)$ induces a distribution $\mu \in \Delta(A_1 \times A_2)$, given by $\mu(a_1,a_2)=\mu_1(a_1)\mu_2(a_2|a_1,a_1)$ for all $(a_1,a_2)$. Moreover, it is easy to verify that a distribution $\mu\in \Delta(A_1 \times A_2)$ is a Bayes correlated distribution if and only if the following two constraints are satisfied:
\begin{itemize}
\item[$(i)$] For all $a_1$ such that $\sum_{a_2}\mu(a_1,a_2)>0$, we have
\beq 
 \sum_{a_2}u_1(a_1,a_2)\mu(a_2|a_1) \geq \max_{ a_1 \in A_1}\min_{a_2 \in A_2} u_1(a_1,a_2).
\eeq 
\item[$(ii)$] For all $a_2$ such that $\sum_{a_1}\mu(a_1,a_2)>0$, we have
\beq 
 \sum_{a_1}u_2(a_1,a_2)\mu(a_1|a_2) \geq  \sum_{a_1}u_2(a_1,a'_2)\mu(a_1|a_2),
\eeq
for all $a_2'$.
\end{itemize}

Condition $(i)$ states that if player 1 is recommended to play $a_1$, but plays $a_1' \neq a_1$ instead, the mediator may recommend player $2$ to punish player $1$, i.e., to play $a_2 \in \arg\min_{a'_2 \in A_2} u_1(a'_1,a'_2)$. Consequently,  any recommendation made to player $1$, which gives player $1$ a payoff higher than his (pure) maxmin payoff, can be sustained as a Bayes correlated equilibrium. Condition $(ii)$ states that all recommendations the mediator makes to player $2$ must be best responses to player 2's belief about player $1$'s action.

For a concrete example, let us revisit Example 1. The strategic-form game is (player 1 is the row player): 
\begin{table}[h!]
\center
\begin{tabular}{|c|c|c|}\hline
& $L$ & $R$ \\ \hline
$T$ & $2,2$ & $0,1$ \\\hline
$B$ & $3,0$ & $1,1$ \\ \hline 
\end{tabular}
\end{table}

The set of Bayes correlated distributions is given by \[\{\mu: \mu(T,L) \geq \mu(B,L), \mu(B,R) \geq \mu(T,R), \mu(T,L) \geq \mu(T,R)\}.\]
Indeed, if $L$ (resp., $R$) is recommended to player 2,  player $2$ must conjecture that player 1 played $T$ (resp., $B$) with probability at least $1/2$ for $L$ (resp., $R$)  to be a best response. We therefore need $\mu(T,L) \geq \mu(B,L)$ and $\mu(B,R) \geq \mu(T,R)$. Moreover, the maxmin payoff to player $1$ is 1. Therefore, if action $T$ is recommended to player $1$, it must be that $2 \mu(T,L)/(\mu(T,L)+\mu(T,R)) \geq 1$, i.e., 
$\mu(T,L) \geq \mu(T,R)$. The associated payoffs are depicted in the figure below (the dark gray triangle):

\begin{figure}[h]
\begin{center}
\begin{tikzpicture}[scale=1.1]
    % Draw axes
    \draw [<->,thick] (0,3) node (yaxis) [above] {$u_2$}
        |- (4,0) node (xaxis) [right] {$u_1$};
    % Draw two intersecting lines
    \draw[fill=lightgray] (3,0)--(2,2)  -- (0,1)--(3,0) ;
    \draw[fill=darkgray] (2.5,1)--(2,2)  -- (1,1)--(2.5,1) ;
    \node[below] at (0,0) {$0$};
     \node[below] at (1,0) {$1$}; 
     \node[below] at (2,0) {$2$};
       \node[below] at (3,0) {$3$};
    \node[left] at (0,1) {$1$};
    \node[left] at (0,2) {$2$};
    \draw[fill] (2.5,1) circle [radius=0.05];
    %\node[left] at (0,3) {$3$};
    \draw[fill] (0,1) circle [radius=0.05];
     \draw[fill] (3,0) circle [radius=0.05];
     \draw[fill](1,1) circle [radius=0.05];
      \draw[fill] (2,2) circle [radius=0.05];
      \end{tikzpicture}
 \end{center}
\caption{Feasible payoffs (light grey) and Bayes correlated equilibrium payoffs (dark grey)}
\end{figure}
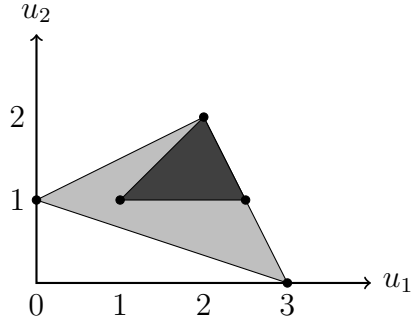

For instance, the payoff $(5/2,1)$ corresponds to the following signaling structure and equilibrium strategies. There are two equally likely signals $t$ and $b$ at the first stage; player $1$ is privately told the first-stage signal. There are two signals at the second-stage $l$ and $r$; player $2$ is privately told the second-stage signal.
Player $2$ receives $l$ if and only if $(T,t)$ and $(B,b)$ are the first-stage profiles of signal and action. An equilibrium of that extended game consists in players playing according to their signals. This gives us the distribution $\mu(T,L)=\mu(B,L)=1/2$ and its associated payoff $(5/2,1)$, as required. 

Finally, note that if we apply the definition of BM to the strategic-form of the game, $\mu(B,R)=1$ is the unique outcome. Indeed, if the mediator recommends a strategy to both players as a function of the  realized signals and states, we simply obtain the correlated equilibria of the game, since there are no signals and states (and the strategies are the actions). This is also the unique distribution induced by the communication equilibria of the game.\footnote{This is also the unique distribution induced by the extensive-form correlated equilibria of the game, as defined by von Stengel and Forges (2008).} To see this, note that it is never optimal for player 1 to play $T$ when recommended to do so. Player 1 can disobey and play $B$, and report to have played $T$ to the mediator.\hfill $\blacksquare$\medskip

\subsection{Additional Equivalence Theorems}

The objective of this section is to enrich our analysis by requiring rational behavior both on and off the equilibrium path.  The main message is that Theorem \ref{th:equiv} generalizes to stronger solution concepts.  \medskip

\subsubsection{Weak Perfect Bayesian Equilibrium} Throughout, we fix an expansion $\Gamma_{\pi}$ of $\Gamma$. We denote $\overline{\proba}_{\sigma,\pi}(\cdot|h^t,m^t,\omega^t)$ the distribution over $HM \Omega$  induced by the profile of behavioral strategies $\sigma$ and the expansion $\pi$, given the history $(h^t,m^t,\omega^t)$. The distribution  $\overline{\proba}_{\sigma,\pi}(\cdot|h^t,m^t,\omega^t)$ is well-defined even if $(h^t,m^t,\omega^t)$ has zero probability under $\proba_{\sigma,\pi}$, and it is equal to $\proba_{\sigma,\pi}(\cdot|h^t,m^t,\omega^t)$ when $\proba_{\sigma,\pi}(h^t,m^t,\omega^t)>0$. Intuitively, this distribution represents the beliefs an outside observer has at $(h^t,m^t,\omega^t)$ if it is conjectured that players continue to follow their equilibrium strategies even after deviations.  We adopt the convention that $\overline{\proba}_{\sigma,\pi}(\bm{h},\bm{m},\bm{\omega}):=\overline{\proba}_{\sigma,\pi}(\bm{h},\bm{m},\bm{\omega}|h^0,m^0,\omega^0)$. At any given history $(h^t,m^t,\omega^t)$, player $i$'s expected payoff  is
\[U_i(\sigma|h^t,m^t,\omega^t):=\sum_{\bm{h},\bm{m},\bm{\omega}}u_i(\bm{h},\bm{\omega})\overline{\proba}_{\sigma,\pi}(\bm{h},\bm{m},\bm{\omega}|h^t,m^t,\omega^t). \]

To complete the description, we need to specify the belief player $i$ has at any private history $(h_i^t,m_i^t)$.  To do so, we specify a belief system $\beta$. Player $i$ believes that the history is $(h^t,m^t,\omega^t)$  with probability  $\beta(h^t,m^t,\omega^t |h^t_i,m_i^t)$ at the private history $(h^t_i,m_i^t)$. At the private history  $(h_i^t,m_i^t)$, player $i$'s expected payoff is therefore:
\beq
U_i(\sigma,\beta|h_i^t,m_i^t):=
 \sum_{\bm{h^t},\bm{m^t},\bm{\omega^t}}U_i(\sigma|\bm{h^t},\bm{m^t},\bm{\omega^t})\beta(\bm{h^t},\bm{m^t},\bm{\omega^t}|h_i^t,m_i^t).\eeq
\medskip 

A profile $\sigma$ of behavioral strategies is a \emph{weak perfect Bayesian equilibrium} of $\Gamma_{\pi}$ if there exists a  belief system $\beta$ on $H\Omega M$ such that: 
 \begin{itemize}
\item[(i)] \textit{Sequential rationality:} For all $t$, for all $i$, for all $(h_i^t,m_i^t)$, 
 \[U_i(\sigma,\beta|h_i^t,m_i^t) \geq U_i((\sigma'_i,\sigma_{-i}),\beta|h_i^t,m_i^t),\]
for all $\sigma'_i$.
 \item[(ii)]  \textit{Belief consistency:} The belief system $\beta$ is consistent with $\sigma$, that is, for all $(h,m,\omega) \in HM\Omega$, for all 
 $(i,t)$, 
\[\beta(h^t,m^t,\omega^t|h_i^t,m_i^t)= \frac{\proba_{\sigma,\pi}(h^t,m^t,\omega^t)}{\proba_{\sigma,\pi}(h_i^t,m_i^t)},
\]
whenever $\proba_{\sigma,\pi}(h_i^t,m_i^t)>0$.
\end{itemize}

We let $w\mathcal{PBE}(\Gamma_{\pi})$ be the set of distributions over $H\Omega$ induced by the weak perfect Bayes equilibria of $\Gamma_{\pi}$. As before, the objective is to provide a characterization of the sets $ \bigcup_{\Gamma_{\pi} \text{\,an  expansion of\;\,} \Gamma}w\mathcal{PBE}(\Gamma_{\pi})$, i.e., we want to characterize the distributions over the outcomes $H\Omega$ of the base game $\Gamma$ that we can induce by means of some expansion $\Gamma_{\pi}$ of the base game, \emph{without} any reference to particular expansions. To do so, we need to introduce the concept of weak perfect Bayes correlated equilibrium of $\Gamma$.

\medskip

\textit{\textbf{Weal Perfect Bayes Correlated Equilibrium.}}  We consider mediated extensions $\mathcal{M}(\Gamma)$ of the game $\Gamma$, where at each stage the set of recommendations made to a player may be a strict subset of the set of actions available to the player. Formally, 
for each private history $(h^t_i,\hat{a}_i^{t-1})$ of past and current signals $s_i^t$, past actions $a_i^{t-1}$ and past recommendations $\hat{a}_i^{t-1}$, $R_{i,t}(h^t_i,\hat{a}_i^{t-1}) \subseteq A_{i,t}$ is the set of possible recommendations to player $i$. We refer to the function 
$R_{i,t}$ as the \emph{mediation range} of player $i$ at stage $t$. We denote $\mathscr{H}(R)$  the set of all terminal histories consistent with the mediation ranges in the mediated extension $\mathcal{M}(\Gamma)$, i.e., 
$(h,\omega,\hat{a}) \in \mathscr{H}(R)$ if and only if $(h,\omega) \in H\Omega$ and $\hat{a}_{i,t} \in R_{i,t}(h^t_i,\hat{a}_i^{t-1})$ for all $i$, for all $t$.

We denote $\overline{\proba}_{\mu\circ\tau,p}(\cdot|h^t,\omega^t,\hat{a}^t)$ the distribution over $\mathscr{H}(R)$ induced by the profile of strategies $\tau$,  the recommendation kernels $\mu$ and the kernels $p$, given the history  $(h^t,\omega^t,\hat{a}^t)$.  At any history  $(h^t,\omega^t,\hat{a}^t)$, player $i$'s expected payoff is
\[U_i(\mu \circ\tau| h^t,\omega^t,\hat{a}^t):=\sum_{\bm{h},\bm{\omega},\bm{\hat{a}}}u_i(\bm{h},\bm{\omega})\overline{\proba}_{\mu\circ\tau,p}(\bm{h},\bm{\omega},\bm{\hat{a}}|h^t,\omega^t,\hat{a}^t).\]

\noindent Finally, at any private history $(h_i^t,\hat{a}_i^t)$, player $i$'s expected payoff is:
\beq
U_i(\mu \circ\tau, \beta| h_i^t,\hat{a}_i^t):= 
\sum_{\bm{h^t},\bm{\omega^t},\bm{\hat{a}^t}}U_i(\mu \circ\tau| \bm{h^t},\bm{\omega^t},\bm{\hat{a}^t})
 \beta( \bm{h^t},\bm{\omega^t},\bm{\hat{a}^t}|h_i^t,\hat{a}_i^t),
\eeq
where $\beta$ is a belief system. We write $\mathcal{T}_i^{*,t}$ for the subset of action strategies of player $i$, where player $i$ is obedient up to (including) stage $t$. We are now ready to define the concept of weak perfect Bayes correlated equilibrium.\medskip

A \textit{weak perfect Bayes correlated equilibrium} of $\Gamma$ is a collection of mediation ranges $R_{i,t}: H_i^{t} \times A_i^{t-1}\rightarrow 2^{A_{i,t}}\setminus \{\emptyset\}$ for all $(i,t)$, a collection of  recommendation kernels $\mu_t(h^t,\omega^t,\hat{a}^{t-1}) : \times_{i \in I}R_{i,t}(h_i^t,\hat{a}^{t-1}_i) \rightarrow [0,1]$, where $\sum_{\hat{a}_t \in \times_{i \in I}R_{i,t}(h_i^t,\hat{a}^{t-1}_i)}\mu_t(h^t,\omega^t,\hat{a}^{t-1})[\hat{a}_t]=1$, for all $(h^t,\omega^t,\hat{a}^{t-1})$ in $\mathscr{H}(R)$ and a belief system $\beta$ such that:
 \begin{itemize}
\item[(i)] \textit{Obedience:} For all $t$, for all $i$, for all private histories $(h_i^t,\hat{a}_i^{t}) $  such that  $\hat{a}_{i,t'} \in R_{i,t'}(h^{t'}_i,\hat{a}_i^{t'-1}) $ for all $t'\leq t$,
\[U_i(\mu \circ \tau^*, \beta| h_i^t,\hat{a}_i^t) \geq U_i(\mu \circ (\tau_{i},\tau^*_{-i}), \beta| h_i^t,\hat{a}_i^t), \]
for all $\tau_i \in  \mathcal{T}_i^{*,t-1}$.
 \item[(ii)]  \textit{Belief consistency:} $\beta$ is consistent with $(\tau^*,\mu,p)$, that is, 
for all $(h,\omega,\hat{a}) \in \mathscr{H}(R)$, for all $(i,t)$,
\[\beta(h^t,\omega^t,\hat{a}^t|h_i^t,\hat{a}_i^t)= \frac{\proba_{\mu\circ \tau^*,p}(h^t,\omega^t,\hat{a}^t)}{\proba_{\mu\circ \tau^*,p}(h_i^t,\hat{a}_i^t)}, \]
whenever $\proba_{\mu\circ \tau^*,p}(h_i^t,\hat{a}_i^t)>0$.
\end{itemize}

 It is worth pausing over the role of the mediation ranges. A weak perfect Bayes correlated equilibrium constrains  the mediator to only recommend actions consistent with the mediation ranges, that is, the only recommendations the mediator can make to player $i$ are in $R_{i,t}(h^t_i,\hat{a}_i^{t-1})$ at history $(h^t,\omega^t,\hat{a}^{t-1})$.\footnote{This explains why the domain of $\mu_t(h^t,\omega^t,\hat{a}^{t-1})$ is $\times_{i \in I}R_{i,t}(h_i^t,\hat{a}^{t-1}_i)$ in our definition.} In addition, players must have an incentive to be obedient at all histories consistent with the mediation ranges. The role of mediation ranges is precisely to insure that players can be obedient at all histories of the mediated game. Without constraining the recommendations the mediator can make, it wouldn't be possible to insure that players are obedient at all histories. E.g., no player would ever have an incentive to play a strictly dominated action. An equivalent formulation is to consider weak perfect Bayesian equilibria of the mediated game, where the mediator is omniscient and unconstrained in its recommendations, and players are obedient on path. The drawback of this alternative formulation is that players do not have to be obedient off path and, therefore, requires to explore all possible behaviors off path. The advantage is that no mediation ranges are required.  We let $w\mathcal{PBCE}(\Gamma)$ be the set of distributions over $H \Omega$ induced by the weak perfect Bayes correlated equilibria of $\Gamma$. \medskip

With all these preliminaries done, we can now state our second equivalence result. 

\begin{theorem}\label{th:equiv-wpbe}
We have the following equivalence: 
\[w\mathcal{PBCE}(\Gamma)= \bigcup_{\Gamma_{\pi} \text{\,an  expansion of\;\,} \Gamma}w\mathcal{PBE}(\Gamma_{\pi}).\]
\end{theorem}
\medskip

Theorem  \ref{th:equiv-wpbe} states an equivalence between (i) the set of distributions over actions, base signals and states induced by all weak perfect Bayes correlated equilibria of $\Gamma$, and  (ii)  the set of distributions over actions, base signals and states we can obtain by considering \emph{all} weak perfect Bayesian equilibria of \emph{all}  expansions of $\Gamma$. \medskip

The logic behind Theorem \ref{th:equiv-wpbe} is identical to the the one behind Theorem \ref{th:equiv}. We can replicate any weak perfect Bayesian equilibrium of $\Gamma_{\pi}$ as a weak perfect Bayesian equilibrium of the auxiliary game $\mathcal{M}^*(\Gamma)$ and then invoke the revelation principle, which was recently proved by Sugaya and Wolitzky (2018, Proposition 2.)  \medskip

We conclude with few additional remarks. First, the set $w\mathcal{PBCE}(\Gamma)$ is convex. To see this, take two distributions $\nu$ and $\nu'$ in $w\mathcal{PBCE}(\Gamma)$. It follows from Theorem \ref{th:equiv-wpbe} that there exist two expansions $\Gamma_{\pi}$ and $\Gamma_{\pi'}$ and two associated weak perfect Bayesian equilibria $(\sigma,\beta)$ and $(\sigma',\beta')$, which induce $\nu$ and $\nu'$, respectively. 
Take $\alpha \in [0,1]$ and consider the expansion $\Gamma_{\alpha \pi + (1-\alpha) \pi'}$, where the information structure $\pi$ (resp., $\pi'$) is drawn with probability $\alpha$ (resp., $1-\alpha$) and the players are informed about the draw. If players coordinate on $\sigma$ (resp., $\sigma'$) when the drawn information structure is $\pi$ (resp., $\pi'$), we obtain the distribution $\alpha \nu+ (1-\alpha) \nu'$. From Theorem \ref{th:equiv-wpbe}, it is in $w\mathcal{PBCE}(\Gamma)$.

Second, despite its theoretical shortcomings, we have considered the concept of weak perfect Bayesian equilibrium as our solution concept.\footnote{It is well-known that weak perfect Bayesian equilibria may not be subgame perfect, may rely on ``irrational'' beliefs, and may not satisfy the one-shot deviation principle.} We did so for two two main reasons. First, it is simple and indeed widely used in applications. Second, it generalizes to continuous games, a common assumption in applications. In what follows, we present another solution concept, which alleviates some of the theoretical shortcomings of weak perfect Bayesian equilibrium. However, it comes at a cost: it is ``harder'' to use in applications.  

\subsubsection{Perfect Bayesian Equilibrium}

 An important tool in modeling off-equilibrium path beliefs is the concept of conditional probability systems (henceforth, CPS). Fix a finite non-empty set $\mathcal{X}$. A conditional probability system $\beta$ on $\mathcal{X}$ is  a function from $2^{\mathcal{X}} \times 2^{\mathcal{X}} \setminus \{\emptyset\}$ to $[0,1]$, which satisfies three properties: for all $X,Y,Z$ with $X \subseteq \mathcal{X}$, $Y \subseteq \mathcal{X}$ and $ \emptyset \neq Z \subseteq \mathcal{X}$,
\begin{itemize}
\item[(i)] $\beta(Z|Z) = 1$ and $\beta(\mathcal{X}|Z) = 1$,
\item[(ii)] if $X \cap Y = \emptyset$, then $\beta(X\cup Y |Z)=  \beta(X |Z)+\beta(Y |Z)$, 
\item[(iii)] if $X\subseteq Y\subseteq Z$  and $Y\neq\emptyset$, then $\beta(X|Z)=\beta(X|Y)\beta(Y|Z)$.
\end{itemize}

\noindent Conditional probability systems capture the idea of ``conditional beliefs''  even after zero-probability events. In particular, 
if $\mathcal{X}$ is the set of terminal histories of a game, a conditional probability system induces a belief system, i.e., a belief over histories at each information set of a player.  A conditional probability system also captures the beliefs players have about the strategies and beliefs of others. Finally, using a conditional probability system to represent the players' beliefs imposes that all differences in beliefs come from differences in information. We refer the reader to Myerson (1986) for more on conditional probability systems.\footnote{Myerson shows that for any conditional probability system $\beta$, there exists a sequence of probability measures $\proba^n$ on $\mathcal{X}$ such that (i) $\proba^n(\{x\})>0$ for all $x \in \mathcal{X}$ and (ii) $\beta = \lim_n \proba^n$, that is, $\beta(X|Y)= \lim_n \frac{\proba^n(X \cap Y)}{\proba^n(Y)}$ for all $X$, for all $Y \neq \emptyset$.}\medskip

\textit{\textbf{Conditional probability perfect Bayesian equilibrium (Sugaya and Wolitzky (2020)).}} We first give an informal definition. A conditional probability perfect Bayesian equilibrium is  a profile of strategies and a conditional probability system such that (i) sequential rationality holds given the belief system induced by the conditional probability system and (ii) the conditional probability system is consistent with the profile of strategies and the data of the game. It is a stronger concept than the concept of weak perfect Bayesian equilibrium and a weaker concept than the concept of sequential equilibrium. We now turn to a formal definition. \medskip

In what follows, we use notation, which parallel the one used in previous definitions, and 
thus do not rehash formal definitions. A \textit{conditional probability perfect Bayesian equilibrium} of $\Gamma_\pi$ is a profile $\sigma$ of behavioral strategies and a CPS $\beta$ on $HM\Omega$, which satisfy: 
 \begin{itemize}
\item[(i)] \textit{Sequential rationality:} For all $t$, for all $i$, for all $(h_i^t,m_i^t)$, 
 \[U_i(\sigma,\beta|h_i^t,m_i^t) \geq U_i((\sigma'_i,\sigma_{-i}),\beta|h_i^t,m_i^t),\]
for all $\sigma'_i$.
 \item[(ii)]  \textit{CPS consistency:} The CPS $\beta$ is consistent with $(\sigma,p,\xi)$, that is, for all $(h,m,\omega) \in HM\Omega$, for all 
 $(i,t)$, 
 \beq
 \beta(a_{t}|h^t,m^t,\omega^t)& =& \prod_{i \in I} \sigma_{i,t}(a_{i,t}|h_i^t,m_i^t),\\
\beta(h_{t+1},\omega_{t+1}|a_t,h^t,m^t,\omega^t)& =&  p_{t+1}(h_{t+1},\omega_{t+1}|a_t,h^t,\omega^t),\\
\beta(m_{t+1}|h^{t+1},m^{t},\omega^{t+1})& =& \xi_{t+1}(m_{t+1}|h^{t+1},m^t,\omega^{t+1}).
\eeq
\end{itemize}

Few comments are worth making. First, to ease notation, we have written $\beta(a_{t}|h^t,m^t,\omega^t)$ for 
\begin{eqnarray*}
\begin{split}
\beta\Big(\Big\{(\bm{h},\bm{m},\bm{\omega}) \in HM\Omega: (\bm{a}_t,\bm{h^t},\bm{m^t},\bm{\omega^t})= & (a_t,h^t,m^t,\omega^t)\Big\}\\ 
& \Big | \Big\{(\bm{h},\bm{m},\bm{\omega}) \in HM\Omega: (\bm{h^t},\bm{m^t},\bm{\omega^t})= (h^t,m^t,\omega^t)\Big\}\Big).
\end{split}
\end{eqnarray*}
We use similar abuse of notation throughout; this should not create any confusion. Second, the consistency of the CPS implies that 
\[\beta(h^t,m^t,\omega^t|h_i^t,m_i^t)= \frac{\proba_{\sigma,\pi}(h^t,m^t,\omega^t)}{\proba_{\sigma,\pi}(h_i^t,m_i^t)},
\]
whenever $\proba_{\sigma,\pi}(h_i^t,m_i^t)>0$. Third, a conditional probability perfect Bayesian equilibrium is subgame perfect. Fourth, since the belief a player has is induced by the CPS, two players with the same information have the same belief.  However, the CPS does not impose a ``don't signal what you don't know''  condition. To do so, we would need to require the CPS to maintain the relative likelihood of any two histories before and after players taking actions. 

\medskip 

   We let $\mathcal{CPPBE}(\Gamma_{\pi})$ be the set of distributions over $H \Omega$ induced by the conditional probability perfect Bayesian equilibria of $\Gamma_{\pi}$. \medskip

As before, the objective is to provide a characterization of the set $ \bigcup_{\Gamma_{\pi} \text{\,an  expansion of\;\,} \Gamma}\mathcal{CPPBE}(\Gamma_{\pi})$, i.e., we want to characterize the distributions over the outcomes $H\Omega$ of the base game $\Gamma$ that we can induce by means of some expansion $\Gamma_{\pi}$ of the base game, \emph{without} any reference to particular expansions. To do so, we need to introduce the concept of sequential Bayes correlated equilibrium of $\Gamma$.

\medskip

\textit{\textbf{Sequential Bayes correlated equilibrium.}} As in the previous section, we consider mediated extensions $\mathcal{M}(\Gamma)$ of the game $\Gamma$, where at each stage the set of recommendations made to a player may be a strict subset of the set of actions available to the player. We use the same notation and do not rehash them. \medskip 

A feedback rule $f:=(f_1,\dots,f_T)$ is a deterministic recommendation kernel, which recommends the action $f_t(h^t,\omega^t)$ at history $(h^t,\omega^t) \in H^t\Omega^t$.  Note that given $f$, the history $(h^t,\omega^t)$ encodes the profile of recommendations $\hat{a}^t$ as $(f_1(h^1,\omega^1),f_2(h^2,\omega^2),\dots, f_t(h^t,\omega^t))$. A feedback rule $f$ is consistent with the mediation ranges $R$ if $f_{i,t}(h^t,\omega^t) \in R_{i,t}(h_i^t,\hat{a}_i^{t-1})$ for all $i$, for all $(h^t,\omega^t)$, for all $t$, where $\hat{a}^{t-1}$ is the profile of recommendations encoded by $f$ at $(h^{t-1},\omega^{t-1})$. We let $\mathcal{F}$ be the set of feedback rules and $\mathcal{F}(R)$ the subset of feedback rules consistent with the mediation ranges $R$.\medskip

We denote $\overline{\proba}_{f \circ \tau,p}(\cdot|h^t,\omega^t)$ the distribution over $\mathscr{H}(R)$ induced by the profile of strategies $\overline{\tau}$,  the feedback rule $f$ and the kernels $p$, given the history  $(h^t,\omega^t)$.  At any history  $(h^t,\omega^t)$, player $i$'s expected payoff is
\[U_i(f \circ \tau| h^t,\omega^t):=\sum_{\bm{h},\bm{\omega}}u_i(\bm{h},\bm{\omega})\overline{\proba}_{f \circ \tau,p}(\bm{h},\bm{\omega}|h^t,\omega^t),\]
when the feedback rule is $f$. Finally, at any private history $(h_i^t,\hat{a}_i^{t})$, player $i$'s expected payoff is:
\beq
U_i(\tau, \beta| h_i^t,\hat{a}_i^t):= 
\sum_{\bm{h^t},\bm{\omega^t},f}U_i(f \circ \tau| h^t,\omega^t)\beta(f,h^t,\omega^t|h_i^t,\hat{a}_i^t),
\eeq
where $\beta$ is a CPS on $\mathcal{F}(R) \times H\Omega$. Here, we write  $\beta(f,h^t,\omega^t|h_i^t,\hat{a}_i^t)$ for: 
\beq
\beta\Big(\Big\{(\bm{f},\bm{h},\bm{\omega}): ( \bm{f},\bm{h}^t,\bm{\omega}^t=f,h^t,\omega^t)\Big\}
\Big|\Big\{(\bm{f},\bm{h},\bm{\omega}):  (\bm{f}_{i,1}(\bm{h}^1,\bm{\omega}^1),\dots, \bm{f}_{i,t}(\bm{h}^t,\bm{\omega}^t))=\hat{a}^t_{i},\bm{h}_i^t=h_i^t\Big\}\Big) \eeq

A communication mechanism $\mu \in \Delta(\mathcal{F})$ is a sequential Bayes correlated equilibrium if there exist mediation ranges $R$ and a conditional probability system $\beta$ on $\mathcal{F}(R) \times H\Omega$ such that: 
 \begin{itemize}
\item[(i)] \textit{Obedience:} For all $t$, for all $i$, for all private histories $(h_i^t,\hat{a}_i^{t}) $  such that  $\hat{a}_{i,t'} \in R_{i,t'}(h^{t'}_i,\hat{a}_i^{t'-1}) $ for all $t'\leq t$,
\[U_i(\tau^*, \beta| h_i^t,a_i^t) \geq U_i((\tau_i,\tau^*_{-i}), \beta| h_i^t,a_i^t) \]
for all $\tau_i \in  \mathcal{T}_i^{*,t-1}$.
 \item[(ii)]  \textit{CPS consistency:} For all $f,h,\omega, t$, 
\begin{flalign*}
& \beta(f,h,\omega)=\mu(f)\overline{\proba}_{f \circ \tau^*,p}(h,\omega) \\
& \beta(f,h,\omega| (f_1,\dots,f_t), (h^t,\omega^t))= \beta(f|(f_1,\dots,f_t), (h^t,\omega^t))\overline{\proba}_{f \circ \tau^*,p}(h,\omega|h^t,\omega^t).
\end{flalign*}
\end{itemize}
Few remarks are worth making. First, in a sequential Bayes correlated equilibrium, players have an incentive to be obedient at all histories consistent with the mediation ranges. Second, unlike previous definitions, the definition asserts that the omniscient mediator selects a feedback rule $f$ with probability $\mu$, i.e., as if the mediator chooses a mixed strategy (and not a behavioral strategy). In addition, the conditional probability system is required to be consistent with $\mu$. We may wonder whether an equivalent formulation exists where the mediator chooses recommendation kernels $(\mu_t)_t$ (behavioral strategies) and consistency is imposed with respect to $(\mu_t)_t$, as we did in the definition of a weak perfect Bayes correlated equilibrium. As Sugaya and Wolitzky (2020) show, the answer is unfortunately no. Intuitively, the current formulation allows more flexibility in choosing beliefs, which is needed for a revelation principle to hold. Third, sequential Bayes correlated equilibria are sequential communication equilibria (Myerson, 1986) of mediated games, where the mediator is omniscient.\footnote{Sequential Bayes correlated equilibria are the subsets of Bayes correlated equilibria, where the mediator never recommends co-dominated actions, a generalization of the concept of dominance.  We refer the reader to Myerson (1986) for more details.} We let $\mathcal{SBCE}(\Gamma)$ be the set of distributions over $H \Omega$ induced by the sequential Bayes correlated equilibria of $\Gamma$.

\begin{theorem}\label{th:equiv-seq}
We have the following equivalence: 
\[\mathcal{SBCE}(\Gamma)=\bigcup_{\Gamma_{\pi} \text{\,an  expansion of\;\,} \Gamma}\mathcal{CPPBE}(\Gamma_{\pi}).\]
\end{theorem}
\medskip

Theorem  \ref{th:equiv-seq} states an equivalence between (i) the set of distributions over actions, base signals and states induced by all sequential Bayes correlated equilibria of $\Gamma$, and  (ii) the set of distributions over actions, base signals and states we can obtain by considering \emph{all} conditional probability perfect Bayesian equilibria of \emph{all}  expansions of $\Gamma$. The logic behind Theorem  \ref{th:equiv-seq} and proof are the same as in previous sections. The set $\mathcal{SBCE}(\Gamma)$ is convex.

\section{Applications}
This section presents two simple applications, which are suggestive of the usefulness of our characterization results. 

\subsection{Rationalizing dynamic choices}
Suppose that an analyst observes the choices of a decision-maker over a finite number of periods, but does not observes the information the decision-maker had. Suppose, furthermore, that the analyst assumes that the state does not change over time.  Which profiles of choices can be rationalized?

de Oliveira and Lamba (2019) have recently addressed that question. These authors assume, however, that the information the decision-maker receives over time is independent of his past actions.\footnote{With our notation, this is equivalen to requiring that $\xi_t(\cdot|(a^{t-1},s^t),m^{t-1},\omega^{t})=\xi_t(\cdot|(\tilde{a}^{t-1},s^t),m^{t-1},\omega^{t})$ for all $(a^{t-1},\tilde{a}^{t-1})$, for all $(s^t,m^{t-1},\omega^{t})$.} Thanks to Theorem \ref{th:equiv}, we are able to generalize their result with little difficulty. Throughout, we follow the terminology of de Oliveira and Lamba.\medskip

We say that the profile of choices $(a_1^*,\dots,a_T^*)$ is rationalizable if there exist a probability $p \in \Delta(\Omega)$, sets of signals $M_t$ and kernels $\xi_t: A^{t-1} \times M^{t-1} \times \Omega \rightarrow \Delta(M_t)$ such that the decision-maker chooses optimally and $(a^*_1,\dots,a_T^*)$ is optimal for some realizations $(\omega,m)$ of states and signals. We assume the decision-maker payoff function $u$ is known to the analyst.\medskip

From Theorem \ref{th:equiv}, the profile of choices $(a^*_1,\dots,a^*_T)$ is \emph{rationalizable} if there exists a probability $p \in \Delta(\Omega)$ and a Bayes correlated equilibrium $\mu$ such that $ \mathbb{P}_{\mu \circ \tau^*,p}(a^*)>0$. Recall that $\mu$ is a Bayes correlation equilibrium if:
\[\sum_{a,\hat{a},\omega} u(a,\omega)\mathbb{P}_{\mu \circ \tau^*,p}(a,\hat{a},\omega) \geq \sum_{a,\hat{a},\omega} u(a,\omega)\mathbb{P}_{\mu \circ\tau,p}(a,\hat{a},\omega), \]  
for all $\tau$. The objective is to derive conditions on the primitives, which guarantee the existence of such a Bayes correlated equilibrium.\medskip 

We say that $D: A \rightarrow \Delta(A)$ is a \emph{deviation plan} if there exists $\tau$ such that
\begin{flalign*}
& D(a_1,\dots,a_T|\hat{a}_1,\dots,\hat{a}_T) :=  \prod_{t=1}^{T}\tau_{t}(a_{t}|\underbrace{(\hat{a}_1,\dots, \hat{a}_{t})}_{\text{recommendations}}, \underbrace{(a_1,\dots,a_{t-1})}_{\text{choices}})
\end{flalign*}
for all $(\hat{a},a)$. A deviation plan specifies what the decision-maker would do if he were to face a fixed sequence of recommendations.\medskip 

\begin{definition}
The profile $a^*$ is \emph{surely} dominated if there exists a deviation plan $D$ such that for all $\omega$, for all $a$, for all $a'$: 
\begin{flalign*}
 u(a^*, \omega) <   & \sum_{t=1}^{T} \sum_{b \in B^t_{a^*}} u(b, \omega)D(b|a^*_1,\dots, a_{t}^*,a'_{t+1},\dots, a'_T),\\
 u(a, \omega) \leq &   \sum_{t=1}^{T} \sum_{b \in B^t_{a}} u(b, \omega)D(b|a_1,\dots, a_{t},a'_{t+1},\dots, a'_T),
\end{flalign*}
where $B^1_a:=(A_1\setminus\{a_1\}) \times A_2 \times \dots A_T$, $B^t_a:= \{(a_1,\dots,a_{t-1})\} \times( A_t \setminus \{a_t\}) \times A_{t+1} \times \dots A_{T}$ for all $t \in \{2,\dots,T-1\}$ and $B^T_a=\{a\}$.
\end{definition}

The set $B_a^t$ is the set of all profiles of choices, which coincide with $a$ up to period $t$ and differ from $a$ at period $t$. Note that $\bigcup_{t=1}^{T} B^t_a=A$ for all $a$. Intuitively, a profile of choices is surely dominated if the decision-maker has a deviation plan which guarantees an improvement regardless of the state $\omega$, the period $t$ at which the decision-maker is first disobedient, \emph{and} the subsequent recommendations $(a'_{t+1},\dots,a'_T)$.   We have the following characterization. 

\begin{theorem}\label{th:rat}
The profile of choices $a^*$ is rationalizable if and only it is not surely dominated. 
\end{theorem}

To understand Theorem \ref{th:rat}, we first rewrite the obedience constraint. Let $f=(f_1,\dots,f_T)$ be a feedback rule, i.e., $f_t: A^{t-1} \times A^{t-1} \times \Omega \rightarrow A$. A feedback rule specifies a deterministic recommendation at each history of past actions, recommendations and states. Note that we voluntarily include past recommendations in the definition of a feedback rule to stress that it is a \emph{pure} strategy.\footnote{Naturally, as we did earlier, we could restrict attention to the histories, which are not excluded by the feedback rule.} Let $\mathcal{F}$ be the finite set of all feedback rules and $\mathcal{F}^*$ be non-empty subset of feedback rules, which recommend $a^*$ on path.

Similarly, we associate a pure strategy $\tau$ with an action rule $g=(g_1,\dots,g_T)$, with $g_t: A^{t-1} \times A^{t-1} \times A \rightarrow A$. The action rule $g$ specifies a deterministic pure action at each history of past actions and past and current recommendations. We associate $\tau^*$ with the rule $g^*$, where $g_t^*(a^{t-1},\hat{a}^{t-1},\hat{a})=\hat{a}$. Let $\mathcal{G}$ be the set of action rules. Thanks to Kuhn's theorem, we can rewrite the condition for rationalization as: there exists $\mu \in \Delta(\mathcal{F} \times \Omega)$ such that $\mu(\mathcal{F}^*)>0$ and
\begin{eqnarray*}
\sum_{f,\omega,g}\sum_{a,\hat{a}} u(a,\omega)\Big(\mathbb{P}_{f,\omega, g^*}(a,\hat{a}) -\mathbb{P}_{f,\omega,g}(a,\hat{a})\Big)\mu(f,\omega)\nu(g) \geq 0,
\end{eqnarray*}
for all $\nu \in \Delta(\mathcal{G})$, where $\mathbb{P}_{f,\omega,g}$ is the degenerate distribution over actions and recommendations induced by the feedback rule $f$ and the action rule $g$ when the state is $\omega$.\footnote{E.g., $\hat{a}=(f_1(\omega),f_2(f_1(\omega),g_1(f_1(\omega)),\omega),...)$ and $a=   (g_1(f_1(\omega)), g_2(f_2(f_1(\omega),g_1(f_1(\omega))),\omega),...)$} With this rewriting, it is clear that if a profile of choices is surely dominated, then it is not rationalizable. Indeed, sure dominance implies that regardless of $f$ and $\omega$, the decision-maker is better off following the deviation plan $D$ than being obedient. More precisely, since the deviation plan $D$ is induced by a behavioral strategy $\tau$, Kuhn's theorem states that there exists an outcome-equivalent mixed strategy $\nu$, which is a profitable deviation from obedience. \medskip
 
 As for necessity, suppose that $a^*$ is not rationalizable. For all $\mu$ such that $\mu(\mathcal{F}^*)>0$, there exists $\nu$ such the obedience constraint is violated, i.e., 
\begin{eqnarray*}
\sup_{\mu: \mu(\mathcal{F}^*)>0} \min_{\nu} \sum_{f,\omega, g}\sum_{a,\hat{a}} u(a,\omega)\Big(\mathbb{P}_{f,\omega,g^*}(a,\hat{a}) -\mathbb{P}_{f,\omega, g}(a,\hat{a})\Big)\mu(f,\omega)\nu(g)<0. 
\end{eqnarray*}
Since the set of $\mu$ such that $\mu(\mathcal{F}^*)>0$ is non-empty and convex (but not compact) and the objective is bi-linear in $(\mu,\nu)$, we can apply Proposition I.1.3 from Mertens, Sorin and Zamir, (2015, p. 6)  to obtain:
\begin{eqnarray*}
\min_{\nu}\sup_{\mu: \mu(\mathcal{F}^*)>0}  \sum_{f,\omega, g}\sum_{a,\hat{a}} u(b,\omega)\Big(\mathbb{P}_{f,\omega,g^*}(a,\hat{a}) -\mathbb{P}_{f,\omega, g}(a,\hat{a})\Big)\mu(f,\omega)\nu(g)<0. 
\end{eqnarray*}
Hence, there exists $\overline{\nu}$ such that for all $\mu$ with $\mu(\mathcal{F}^*)>0$,
\begin{eqnarray*}
\sum_{f,\omega, g}\sum_{a,\hat{a}} u(b,\omega)\Big(\mathbb{P}_{f,\omega,g^*}(a,\hat{a}) -\mathbb{P}_{f,\omega, g}(a,\hat{a})\Big)\mu(f,\omega)\overline{\nu}(g)<0. 
\end{eqnarray*}
The result follows then immediately by constructing the behavioral strategy $\overline{\tau}$ induced by $\overline{\nu}$ and its associated deviation plan $\overline{D}$ and considering all $(f,\omega)$.\medskip

As already alluded to, Theorem \ref{th:rat} generalizes  a recent result by de Oliveira and Lamba (2019). Their main result states that the profile $a^*$ is rationalizable if and only if it is not \emph{truly} dominated, with $a^*$ being truly dominated if there exists a deviation rule $D$ such that:
\begin{eqnarray*}
u(a^*,\omega) < &\sum_{b} u(b,\omega)D(b|a^*),\\
u(a,\omega) \leq & \sum_{b} u(b,\omega)D(b|a),
\end{eqnarray*}
for all $\omega$, for all $a$. \medskip 

Clearly, if a profile $a^*$ is \emph{surely} dominated, then it is \emph{truly} dominated. Indeed, if we choose $(a'_{t+1},\dots,a'_T)$ to be equal to $(a_{t+1},\dots,a_T)$ for all $(a,t)$, then we recover the condition for true dominance. However, the converse is not true as the example in Table \ref{table:rat} demonstrates. There are two states, $\omega$ and $\omega'$, three actions, $\ell(\text{eft}),c(\text{enter}),r(\text{ight})$, and two periods. The inter-temporal payoff is the sum of the per-period payoff in Table  \ref{table:rat}. 

\begin{table}[h]
\caption{$(\ell,c)$ is rationalizable and truly dominated}
\begin{tabular}{|c|c|c|c|} \hline
& $\ell$ & $c$ & $r$ \\ \hline
$\omega$ & $0$ & $1$ & $0$ \\ \hline
$\omega'$ & $0$ & $0$ & $1$ \\ \hline
\end{tabular}\label{table:rat}
\end{table}

We now argue that $(\ell,c)$ is \emph{truly} dominated.  Intuitively, since $\ell$ is strictly dominated, the decision-maker benefits from playing a mixture of $c$ and $r$ instead of $\ell$ in the first period. More formally, consider the behavioral strategy $\tau$ given by $\tau_1(c|\ell)=\tau_1(r|\ell)=1/2$, $\tau_1(r|r)=\tau_1(c|c)=1$ and $\tau_2=\tau_2^*$.   The induced deviation rule is $D(c\ell|\ell\ell)=D(r\ell|\ell\ell)=D(cc|\ell c)=D(rc|\ell c)=D(cr|\ell r)=D(rr|\ell r)=1/2$ and $D(a_1a_2|\hat{a}_1\hat{a}_2)=1$ for all other profiles $(a_1,a_2)$ and $(\hat{a}_1,\hat{a}_2)$ such that $(a_1,a_2)=(\hat{a}_1,\hat{a}_2)$. It is then easy to verify that $(\ell,c)$ is indeed truly dominated.\medskip

Yet, it is not {surely} dominated and, therefore, is rationalizable. Intuitively, if the decision-maker learns the state after playing $\ell$ in the first period but does not get any additional information otherwise, he has an incentive to play $\ell$. A Bayes correlated equilibrium is as follows:  the mediator recommends $\ell$ at the first period, regardless of the state, and recommends  $c$ (resp., $r$) at the second period   if and only if the decision-maker has been obedient and the state is $\omega$ (resp., $\omega'$).  If the decision-maker disobeys the recommendation, the mediator recommends then either $c$ or $r$, independently of the state. \medskip

To conclude, this application illustrates how we can apply our results to derive testable implications in dynamic decision problems. We stress that our results  apply equally to dynamic games, including games with evolving states, and thus offer a wider scope for applications.

\subsection{Bilateral Bargaining}

We consider a variation on the work of Bergemann, Brooks and Morris (2013). 
There are one buyer and one seller. The seller makes an offer $a_1 \in A_1 \subset \mathbb{R}_+$ to the buyer, who observes the offer and either accepts ($a_2=1$) or rejects ($a_2=0$) it.  If the buyer accepts the offer $a_1$, the payoff to the buyer is $\omega - a_1$, while the payoff to the seller is $a_1$, with $\omega$ being the buyer's valuation (the payoff-relevant state). We assume that $\omega \in \Omega \subset \mathbb{R}_{++}$. If the buyer rejects the offer, the payoff to both the seller and the buyer is normalized to zero. The buyer and the seller are symmetrically informed and believe that the state is $\omega$  with probability $p(\omega) >0$. We assume that the set of offers the seller can make is finite, but as fine as needed. For future reference, we write $\omega_L$ for the lowest state, $\omega_L^{-}$ for the largest offer $a_1$ strictly smaller than $\omega_L$, and $\omega_H$ for the highest state. \medskip

This model differs from Bergemann, Brooks and Morris (2013) in one important aspect. In our model, both the seller and the buyer have no initial private information about the state, while Bergemann, Brooks and Morris assume that the buyer is privately informed of the state $\omega$. The base game  of  Bergemann, Brooks and Morris thus corresponds to a particular expansion of our base game. Similarly, Roesler and Szentes (2017) consider all information structures, where the buyer has some signals about his own valuation (and the seller is uninformed).  Unlike these papers, we consider \emph{all}  information structures. In particular, the information the buyer receives may depend on the information the seller has received as well as the offer made. In addition, the seller can  be better informed than the buyer in our model. \medskip

We characterize the set of \emph{sequential Bayes correlated equilibria}. A communication system $\mu$ is a sequential Bayes correlated equilibrium if there exist mediation ranges $(R_1,R_2)$ and a conditional probability system $\beta$, which jointly satisfy the following constraints. First, if the omniscient mediator recommends $f_1(\omega) \in R_1$ to the seller, the seller must have an incentive to be obedient, i.e.,  
\beq
 \sum_{f,\omega} f_1(\omega) f_2(f_1(\omega),\omega) \beta(f,\omega|f_1(\omega)) \geq \sum_{f,\omega} a_1 f_2(a_1,\omega) \beta(f,\omega|f_1(\omega))
 \eeq
for all $a_1$. Second, if the offer made to the buyer is $a_1$ and the mediator recommends $f_2(a_1,\omega) \in R_2(a_1)$ to the buyer, the buyer must have an incentive to be obedient, i.e., 
\beq
\sum_{f,\omega} (\omega-a_1)f_2(a_1,\omega) \beta(f,\omega|a_1,f_2(a_1,\omega)) \geq \sum_{f,\omega} (\omega-a_1)(1-f_2(a_1,\omega)) \beta(f,\omega|a_1,f_2(a_1,\omega)).
\eeq
Third, the conditional probability system must be consistent, that is, for all $f \in \mathcal{F}(R)$, for all $a_1,a_2,\omega$,
\beq
\beta(f,a_1,a_2,\omega) &= &\mu(f)p(\omega)\indic\{f_1(\omega)=a_1,f_2(f_1(\omega),\omega)=a_2\},\\
\beta(f,a_1,a_2,\omega|f_1,a_1,\omega)&=&  \beta(f|f_1,a_1,\omega)\indic\{(f_2(a_1,\omega),\omega)=a_2\}.
\eeq

\medskip 

There are immediate bounds on the equilibrium payoffs: the sum of the buyer and seller's payoffs is bounded from above by $\mathbb{E}(\omega)=\sum_\omega p(\omega)\omega$, the buyer's payoff is bounded from below by $0$, and the seller's payoff is bounded from below by $\omega_L^{-}$. The following proposition states that there are, in fact, no other restrictions on equilibrium payoffs. 

\begin{proposition}\label{propBB}
The set of sequential Bayes correlated equilibrium payoffs is \[\co\{(0,\omega_L^{-}), (0,\mathbb{E}(\omega)), (\mathbb{E}(\omega)-\omega_L^{-},\omega_L^{-})\}.\]
\end{proposition}

The set of equilibrium payoffs is depicted in the figure below. 

\begin{figure}[h]
\begin{center}
\begin{tikzpicture}[scale=1.1]
    % Draw axes
    \draw [<->,thick] (0,4) node (yaxis) [above] {seller's payoff}
        |- (4,0) node (xaxis) [right] {buyer's payoff};
    % Draw two intersecting lines
    \draw[fill=darkgray] (0,1)--(0,3)  -- (2.5,1)--(0,1) ;
    \node[below] at (2.5,1) {$(\mathbb{E}(\omega)-\omega_{L}^{-}, \omega_L^{-})$};
    \node[left] at (0,3) {$(0,\mathbb{E}(\omega))$};
    \node[below left] at (0,1) {$(0,\omega_L^{-})$};
    \draw[fill] (0,3) circle [radius=0.05];
     \draw[fill] (2.5,1) circle [radius=0.05];
      \draw[fill] (0,1) circle [radius=0.05];
 \end{tikzpicture}
 \end{center}
 \caption{Payoffs at all sequential Bayes correlated equilibria}
 \end{figure}
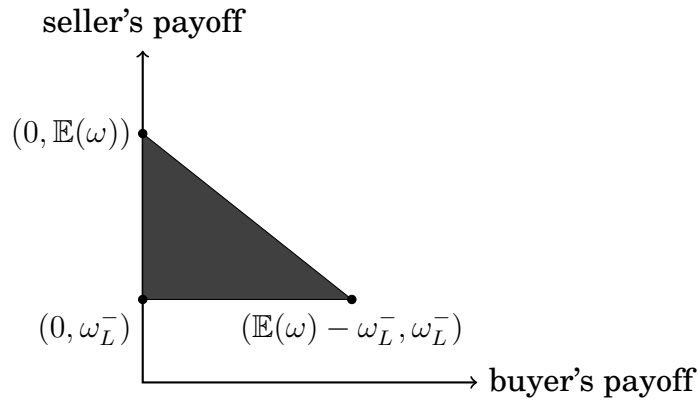
 
 \medskip

\noindent We prove this proposition in what follows. As a preliminary observation, note that the conditional probability system puts no restriction on the buyer's beliefs after observing an off-path offer $a_1$, i.e., an offer such that $\sum_{f,\omega}\mu(f)p(\omega)\indic\{f_1(\omega)=a_1\} =0$. To see this, for any conditional probability system, $\beta(a_1,\omega)=\beta(\omega,a_1|a_1)\beta(a_1)$. Moreover, from the consistency of $\beta$, we have that $\beta(a_1,\omega)=\sum_{f} \beta(f,a_1,\omega)= \sum_{f}\mu(f)p(\omega)\indic\{f_1(\omega)=a_1\}=0$. Since $\beta(a_1)=0$, $\beta(\omega,a_1|a_1)$ is arbitrary and, thus, we can assume that the buyer believes that the state is $\omega_L$ with probability one. We refer to those beliefs as the most pessimistic beliefs. Similarly, there are no restrictions on the buyer's beliefs after observing an off-path offer $a_1$ and a recommendation $f_2(a_1,\omega)$.

We are now ready to state how to obtain the payoff profile $(\mathbb{E}(\omega)-\omega_L^{-},\omega_L^{-})$. We first start with an informal description. The mediator recommends the seller to offer $\omega_L^{-}$, regardless of the state. If the offer $\omega_L^{-}$ is made, the mediator recommends the buyer to accept, regardless of the state. If any offer $a_1 > \omega_L^{-}$ is made, the mediator recommends the buyer to reject the offer, regardless of the state. Since any such offer is off-path, the buyer has an incentive to be obedient when he believes that the state is $\omega_L$ with probability one. As we have just argued, we can choose a well-defined conditional probability system capturing such beliefs. Finally, if any offer $a_1< \omega_L^{-}$  is made, the mediator recommends the buyer to accept, regardless of the state. Formally, the communication system puts probability one to $f$, given by $f_1(\omega)=\omega_{L}^{-}$, $f_2(a_1,\omega)= 0$ if $a_1 >\omega_{L}^{-}$ and $f_2(a_1,\omega)= 1$ if $a_1 \leq \omega_{L}^{-}$ for all $\omega$.  The mediation ranges are $R_1=\{\omega_L^-\}$,  $R_2(a_1)=\{1\}$ if $a_1<\omega_L$, 
$R_2(\omega_L) \subseteq\{0,1\}$, and $R_2(a_1)= \{0\}$ if $a_1 > \omega_L$. \medskip

We now turn our attention to the two other payoff profiles $(0,\mathbb{E}(\omega))$ and $(0,\omega_L^{-})$.  The profile $(0,\mathbb{E}(\omega))$ corresponds to full surplus extraction, which can be obtained with $f_1(\omega)= \omega$ for all $\omega$ and $f_2(a_1,\omega)=1$ whenever $a_1 \leq \omega$ (and zero, otherwise). The mediation ranges are $R_1=\Omega$,  $R_2(a_1)= \{0\}$ if $a_1 > \omega_H$, $R_2(a_1)=\{1\}$ if $a_1<\omega_L$, and $R_2(a_1)=\{0,1\}$ if $a_1 \in \Omega$.\medskip

Lastly, when $\mathbb{E}(\omega) \in A_1$ (which we assume), the profile  $(0,\omega_L^{-})$ is  implementable as follows.  Consider two feedback rules $f$ and $f'$ such that for all $\omega$, $f_1(\omega)=f'_1(\omega)=\mathbb{E}(\omega)$, $f_2(a_1,\omega) = f_2'(a_1,\omega)=0$ if $a_1 > \mathbb{E}(\omega)$, $f_2(a_1,\omega) = f_2'(a_1,\omega)=1$ if $a_1 < \mathbb{E}(\omega)$, $f_2(\mathbb{E}(\omega),\omega)=1$ while $f'_2(\mathbb{E}(\omega),\omega)=0$. Assume that $\mu(f)=\omega_L^{-}/\mathbb{E}(\omega)$, $\mu(f')=1-\mu(f)$, and that $R_1=\{\mathbb{E}(\omega)\}$, $R_2(a_1)=\{1\}$ if $a_1<\omega_L$, $R_2(a_1)=\{0,1\}$ if $a_1=\mathbb{E}(\omega)$, and $R_2(a_1)= \{0\}$, otherwise. In effect, the mediator recommends the seller to offer $\mathbb{E}(\omega)$, regardless of the state, and the buyer to accept that offer with probability $\omega_L^{-}/\mathbb{E}(\omega)$, on path. Off-path, we again use the most pessimistic beliefs to give the seller a payoff of zero, if he deviates. To complete the proof of Proposition \ref{propBB}, it is enough to invoke the bounds on the payoff profiles and the convexity of the set of sequential Bayes correlated equilibrium payoffs.   \medskip

\section{Conclusion}
This paper generalizes the concept of Bayes correlated equilibrium to multi-stage games and offers two applications, which are suggestive of the usefulness of our characterization results. The main contribution is methodological.\medskip 

The reader may wonder why we have not considered the concept of sequential equilibrium. The main reason is that the revelation principle does not hold for this concept. To be more precise, Sugaya and Wolitzky (2020) show that the set of sequential communication equilibria of a multi-stage game characterizes the set of equilibrium distributions we can obtain by considering all mediated extensions of the multi-stage game, where the solution concept is sequential equilibrium. However, their definition of a sequential equilibrium treats the mediator as a player and, thus, allows for the mediator to tremble. When we consider an expansion and its emulation by a mediator with the mediated game $\mathcal{M}^*(\Gamma)$, players do not expect the mediator to tremble. If a player observes an unexpected additional signal, that player must believe with probability one that one of his opponents has deviated. He cannot believe that none of his opponents deviated, but the mediator did. This would be inconsistent with the expansion being the game actually played. Extending the analysis to other solution concepts such as sequential equilibrium or rationalizability or to general extensive-form games is challenging and left for future research.

\appendix
\section*{Appendices}
%\addcontentsline{toc}{section}{Appendices}
\renewcommand{\thesubsection}{\Alph{subsection}}
%\section{Appendix}

\subsection{Proof of Theorem \ref{th:equiv}} 
\medskip

$(\Leftarrow.) $  We first prove that 
$\bigcup_{\Gamma_{\pi} \text{\,an  expansion of\;\,} \Gamma}\mathcal{BNE}(\Gamma_{\pi}) \subseteq \mathcal{BCE}(\Gamma)$.  
Throughout, we fix an  expansion $\Gamma_{\pi}$ of $\Gamma$.  Recall that there exist kernels $(\xi_t)_t$ such that:
\[\pi_{t+1}(h_{t+1},m_{t+1},\omega_{t+1}|a_{t},h^{t},m^{t},\omega^{t})=\xi_{t+1}(m_{t+1} |h^{t+1},m^{t},\omega^{t+1})p_{t+1}(h_{t+1},\omega_{t+1}|a_{t},h^{t},\omega^{t}), \]
for all $(h^{t+1},m^{t+1},\omega^{t+1})$, for all $t$.\medskip

Let $\sigma^*$ be a Bayes-Nash equilibrium of $\Gamma_{\pi}$. We now construct an auxiliary mediated game $\mathcal{M}^*(\Gamma)$, which emulates the distribution $\mathbb{P}_{\sigma^*,\pi}$ as an equilibrium distribution.\medskip 

 The game $\mathcal{M}^*(\Gamma)$ has one additional player, labelled player 0, and a (Forges-Myerson) mediator. Player 0 is a dummy player: his payoff is identically zero. \medskip 
 
 The game unfolds as follows: At stage $t=1$, 
 \begin{itemize}
\item[-] Nature draws $(h_1,\omega_1)$ with probability $p_1(h_1,\omega_1)$.
\item[-] Player $i \in I $ observes the signal $h_{i,1}$ and player $0$ observes $(h_1,\omega_1)$. 
\item[-] Player $0$ reports $(\hat{h}_1,\hat{\omega}_1)$ to the mediator. All other players do not make reports. 
\item[-] The mediator draws the message $m_1$ with probability $\xi_1(m_1|\hat{h}_1,\hat{\omega}_1)$ and sends the message $m_{i,1}$  to player $i$. Player 0 does not receive a message.
\item[-] Player $i$ takes an action  $a_{i,1}$. Player 0 does not take an action.
\end{itemize}
 
Consider now a history $(a_{t-1},h^{t-1},\omega^{t-1})$ of past actions, signals and states and a history $((\hat{h}^{t-1},\hat{\omega}^{t-1}),m^{t-1})$ of reports and messages. At stage $t$:

\begin{itemize}
\item[-] Nature draws $(h_t,\omega_t)$ with probability $p_t(h_t,\omega_t|a_{t-1},h^{t-1},\omega^{t-1})$.
\item[-] Player $i \in I $ observes the signal $h_{i,t}$ and player $0$ observes $(h_t,\omega_t)$. 
\item[-] Player $0$ reports $(\hat{h}_t,\hat{\omega}_t)$ to the mediator. All other players do not make reports. 
\item[-] The mediator draws the message $m_t$ with probability $\xi_t(m_t|\hat{h}^t,m^{t-1}, \hat{\omega}^t)$ and sends the message $m_{i,t}$  to player $i$. Player 0 does not receive a message.
\item[-] Player $i$ takes an action  $a_{i,t}$. Player 0 does not take an action.
\end{itemize}

In the above description, when we say that player $i$ does not make a report, we implicitly assume that the set of reports player $i$ can make to the mediator is a singleton. Similarly, when we sat that player 0 does not take an action. In the rest of the proof, we omit these trivial reports and actions.\medskip

At stage $t$, player $i$'s private history is  therefore $(h_{i}^t,m_i^t)$, which is also player $i$'s private history in $\Gamma_{\pi}$. Thus, $\sigma_i^*$ is a valid strategy for player $i$ in $\mathcal{M}^*(\Gamma)$. Moreover, if player $0$ truthfully reports his private information $(h_t,\omega_t)$ at all histories $((h_t,\omega_t),(h^{t-1},\omega^{t-1}),(\hat{h}^{t-1},\hat{\omega}^{t-1}))$, the conditional probability of the message $m_t$ is the same as in $\Gamma_{\pi}$. It follows immediately that $\sigma^*$ together with the truthful strategy for player $0$ is a Bayes-Nash equilibrium of the auxiliary mediated game $\mathcal{M}^*(\Gamma)$.\medskip 

 From the revelation principle of Forges (1986) and Myerson (1986), there exists a canonical equilibrium $\mu$, where the mediator recommends actions and players are truthful and obedient, provided they have been in the past. At truthful histories, the mediator recommends $\hat{a}_t$ with probability \[\mu_t(\hat{a}_t|\underbrace{(h^t,\omega^t)}_{\text{player 0}},\underbrace{(h_1^t,\dots,h_n^t)}_{\text{players in $I$}}, \underbrace{\hat{a}^{t-1}}_{\text{past recommendations}}).\] 
 
It is then routine to verify that we have a Bayes-correlated equilibrium with the recommendation kernel $\mu_t$ given by
\[\mu_t(\hat{a}_t|h^t,\omega^t,\hat{a}^{t-1}):=\mu_t(\hat{a}_t|(h^t,\omega^t), (h_1^t,\dots,h_n^t), \hat{a}^{t-1}),\]
 for all $(h^t,\omega^t,\hat{a}^{t-1})$ for all $t$.

$(\Rightarrow)$.  We now prove that 
$\mathcal{BCE}(\Gamma)  \subseteq \bigcup_{\Gamma_{\pi} \text{\,an  expansion of\;\,} \Gamma}\mathcal{BNE}(\Gamma_{\pi})$.

Let $\mu$ be a Bayes correlated equilibrium with distribution $\proba_{\mu \circ \tau^*,p}$. We now construct an expansion $\Gamma_{\pi}$ and a Bayes-Nash equilibrium $\sigma^*$ of $\Gamma_{\pi}$, with the property that $\marg_{H\Omega} \proba_{\sigma^* ,\pi}= \marg_{H\Omega} \proba_{\mu \circ \tau^*,p}$.

The expansion is as follows. Let $M_{i,t} =A_{i,t}$ for all $(i,t)$, 
\[\pi_1(h_1,m_1,\omega_1)=p_1(h_1,\omega_1)\mu_1(\hat{a}_1|h_1,\omega_1),\]
 with $m_1 =\hat{a}_1$, for all $(h_1,m_1,\omega_1)$, and 
\[\pi_{t+1}(h_{t+1},m_{t+1},\omega_{t+1}|a_t,h^t,m^t,\omega^t)=p_{t+1}(h_{t+1},\omega_{t+1}|a_t,h^t,\omega^t)\mu_{t+1}(\hat{a}_{t+1}|h^{t+1},\omega^{t+1},\hat{a}^t), \]
with $(m^t,m_{t+1})= (\hat{a}^t,\hat{a}_{t+1})$, for all $(a_t,h^t,m^t,\omega^t, h_{t+1},m_{t+1},\omega_{t+1})$. Clearly, the expansion is well-defined:  $\xi_{1}(m_{1}|h_{1},\omega_{1}) =\mu_{1}(\hat{a}_{1}|h_{1},\omega_{1})$ 
with $m_{1}= \hat{a}_{1}$, and, for $t>1$, $\xi_{t+1}(m_{t+1}|h^{t+1},m^t,\omega^{t+1}) =\mu_{t+1}(\hat{a}_{t+1}|h^{t+1},\omega^{t+1},\hat{a}^t)$ 
with $(m^t,m_{t+1})= (\hat{a}^t,\hat{a}_{t+1})$.

By construction, any strategy $\tau_t : H^t \times A^t \rightarrow \Delta(A_t)$ of $\mathcal{M}(\Gamma)$ is equivalent to a strategy $\sigma_t:H^t \times M^t \rightarrow \Delta(A_t)$ of $\Gamma_{\pi}$, i.e., $\sigma_t(a_t|h^t,m^t):=\times_i \sigma_{i,t}(a_{i,t}|h_i^t,m_i^t)= \times_i \tau_{i,t}(a_{i,t}|h_i^t,\hat{a}_i^t)$ with $m^t=\hat{a}^t$, with the property that $\proba_{\sigma,\pi}(h^t,m^t,\omega^t)= \proba_{\mu \circ \tau,p}(h^t,\hat{a}^t,\omega^t)$ when $m^t=\hat{a}^t$, for all $(h^t,m^t,\omega^t)$, for all $t$. 

To see this last point, note that the definition of $\pi_1$ is clearly equivalent to $\proba_{\sigma,\pi}(h_1,m_1,\omega_1)=\proba_{\mu \circ \tau,p}(h_1,\omega_1,\hat{a}_1)$ with $m_1=\hat{a}_1$, for all $(h_1,m_1,\omega_1)$. By induction, assume that $\proba_{\sigma,\pi}(h^t,m^t,\omega^t)= \proba_{\mu \circ \tau,p}(h^t,\omega^t,\hat{a}^t)$ with $m^t=\hat{a}^t$, for all $(h^t,m^t,\omega^t)$. We now compute the probability of $(h^{t+1},m^{t+1},\omega^{t+1})$.  We have that 
\begin{eqnarray*}
\proba_{\sigma,\pi}(h^{t+1},m^{t+1},\omega^{t+1}) &= & \proba_{\sigma,\pi}(h_{t+1},m_{t+1},\omega_{t+1}|h^{t},m^{t},\omega^{t})\proba_{\sigma,\pi}(h^{t},m^{t},\omega^{t}) \\
& = & \pi_{t+1}(h_{t+1},m_{t+1},\omega_{t+1}|a_t,h^t,m^t,\omega^t)\sigma_t(a_t|h^t,m^t)\proba_{\sigma,\pi}(h^{t},m^{t},\omega^{t}) \\
& = &  p_{t+1}(h_{t+1},\omega_{t+1}|a_t,h^t,\omega^t)\mu_{t+1}(\hat{a}_{t+1}|h^{t+1},\omega^{t+1},\hat{a}^t)\tau_t(a_t|h^t,\hat{a}^t)\proba_{\mu \circ \tau,p}(h^t,\hat{a}^t,\omega^t) \\
& = & \proba_{\mu \circ \tau,p}(h^{t+1},\omega^{t+1},\hat{a}^{t+1}),
\end{eqnarray*}
with $\hat{a}^{t+1} = m^{t+1}$.
Finally, since $\mu$ is a Bayes correlated equilibrium of $\mathcal{M}(\Gamma)$, the strategy $\sigma^* \equiv \tau^*$ is a Bayes-Nash equilibrium of $\Gamma_{\pi}$ and, thus, 
\[\mathcal{BCE}(\Gamma)  \subseteq \bigcup_{\Gamma_{\pi} \text{\,an  expansion of\;\,} \Gamma}\mathcal{BNE}(\Gamma_{\pi}).\]  This completes the proof. 
\subsection{Proof of Theorem \ref{th:equiv-wpbe}} The proof is nearly identical to the proof of Theorem \ref{th:equiv} and is, therefore, omitted. 
We only sketch the minor differences. 

($\Leftarrow$.)  Fix an expansion $\Gamma_{\pi}$ and a weak perfect Bayesian equilibrium $(\sigma^*,\beta)$ of $\Gamma_{\pi}$. We need to construct a weak perfect Bayesian equilibrium of the auxiliary game $\mathcal{M}^*(\Gamma)$, which replicates the distribution $\mathbb{P}_{\sigma, \pi}$. To do so, we define a belief system $\beta^*$ of the auxiliary game $\mathcal{M}^*(\Gamma)$  as follows: $\beta^*(h^t,m^t, \omega^t, (h^t,\omega^t), (h^t,\omega^t)| h_i^t,m_i^t):= \beta(h^t,m^t, \omega^t| h_i^t,m_i^t)$ for all $(h^t,m^t,\omega^t)$, for all $(i,t)$. (In $\mathcal{M}^*(\Gamma)$, player $i$ also has beliefs about the signals $(h^t,\omega^t)$ player 0 receives and the reports $(\hat{h}^t,\hat{\omega}^t)$ by player $0$ to the mediator.) It is immediate to verify that $(\sigma_0^*,\sigma^*,\beta^*)$ is a weak perfect Bayesian equilibrium of $\mathcal{M}^*(\Gamma)$, where $\sigma_0^*$ is the truthful reporting strategy of player 0. The proof then follows from the revelation principle for weak perfect Bayesian equilibrium. The mediation ranges and the belief system come from the revelation principle.\medskip 

($\Rightarrow$.) We construct the expansions as in the  the proof of Theorem \ref{th:equiv}, i.e., defining the additional signals as the recommendations.  Since the additional signals player $i$ can receive are the recommendations, player $i$ can only receive additional signals consistent with the mediation ranges. Thus, we can use the belief system of the weak perfect Bayes correlated equilibrium to construct the weak perfect Bayesian equilibrium of $\Gamma_{\pi}$. 

\subsection{Proof of Theorem \ref{th:equiv-seq}}
The proof is yet again nearly identical to the proof of Theorem \ref{th:equiv}. We only sketch the main differences.

($\Leftarrow$). Fix an expansion $\Gamma_{\pi}$ and a conditional probability perfect Bayesian equilibrium $(\sigma^*,\beta)$ of $\Gamma_{\pi}$. As in the previous proofs, we construct a conditional probability perfect Bayesian equilibrium  of the mediated game $\mathcal{M}^*(\Gamma)$, which replicates the distribution $\mathbb{P}_{\sigma^*,\pi}$. As in the proof of Theorem \ref{th:equiv-wpbe}, we construct a conditional probability system $\beta^*$ of the mediated game $\mathcal{M}^*(\Gamma)$ from the conditional probability system $\beta$ of the game $\Gamma_{\pi}$ such that $((\sigma_0^*,\sigma^*),\beta^*)$ is a conditional probability perfect Bayesian equilibrium  of $\mathcal{M}^*(\Gamma)$, with player 0, the dummy player, truthfully reporting his private information $(h^t,\omega^t)$ at each stage $t$.  Since $\beta$ is a conditional probability system, there exists a sequence $\beta^n$ of fully supported probabilities such that 
$\beta(X |Y)= \lim_{n}\frac{\beta^n(X \cap Y}{\beta^n(Y)}$  for all $X$ and all non-empty $Y$. Consider now the sequence of fully supported kernels $\gamma^n: HM\Omega \rightarrow \Delta(H\Omega \times H\Omega)$, where $\gamma^n$  converges to $\gamma( (h,\omega), (h,\omega)| (h,m,\omega)=1$ for all $(h,m,\omega)$. The interpretation is that player 0 learns and truthfully report $(h,\omega)$, when the state the profile of actions, signals, and states is $(h,m,\omega)$. Let $\beta^*$ be the CPS resulting from taking the limit of $\beta^n \times \gamma^n$. By construction, $\beta^*(h^t,m^t,\omega^t|h_i^t,m_i^t)=\beta(h^t,m^t,\omega^t|h_i^t,m_i^t)$, so that $\sigma^*_i$ remains sequentially rational for player $i$. Moreover, the newly constructed conditional probability system is consistent with the kernels $p$ and $\xi$.   The rest of the proof follows from the revelation principle. 
\medskip

($\Rightarrow$). Let $(\mu,R,\beta)$ be a sequential Bayes correlated equilibrium. The difference with the previous proofs is that the definition of a sequential Bayes correlated equilibrium does not specify recommendation kernels  $(\mu_1,\dots,\mu_T)$, which can then be used as expansions. However, as in the proof of Kuhn's theorem, we can construct such recommendation kernels from $\mu$. \medskip 

The construction is iterative. In the sequel, we slightly abuse notation and write $(\mu_1,\dots,\mu_T)$ for the kernels. For all $(h^1,\omega^1,\hat{a}^0)$ such that $p_1(h_1,\omega_1)>0$,
\[\mu_1(\hat{a}_1|\hat{a}^0,h^1,\omega^1):=\sum_{f} \mu(f)\indic\{f_1(h^1,\omega^1)=\hat{a}_1\}. \]
 Note that $\mu_1(\hat{a}_1|h^1,\omega^1, \hat{a}^0)>0$ and $\sum_{\hat{a}_1}\mu_1(\hat{a}_1|h^1,\omega^1,\hat{a}^0)=1$. The kernel $\mu_1$ is thus well-defined. (Recall that $(h^1,\omega^1)=(h_1,\omega_1)$ and that $\hat{a}^0$ is a singleton.)\medskip 
 
 We proceed iteratively. For all $(h^t,\omega^t,\hat{a}^{t-1})$ such that 
 \beq
 p_1(h_1,\omega_1)\mu_1(\hat{a}_1|\hat{a}^0,h^1,\omega^1) \times \dots \times \\ 
p_{t-1}(h_{t-1},\omega_{t-1}|a_{t-1},h^{t-2},\omega^{t-2})\mu_{t-1}(\hat{a}_{t-1}|\hat{a}^{t-2},h^{t-1},\omega^{t-1}t)p_t(h_t,\omega_t|a_t,h^{t-1},\omega^{t-1})>0
 \eeq
 for some $(a_1,\dots,a_t)$, 
 \[\mu_t(\hat{a}_t|\hat{a}^{t-1},h^t,\omega^t):=\frac{\sum_{f} \mu(f)\indic\{f_1(h^1,\omega^1)=\hat{a}_1,\dots,f_t(h^t,\omega^t)=\hat{a}_t\}}{\sum_{f} \mu(f)\indic\{f_1(h^1,\omega^1)=\hat{a}_1,\dots,f_{t-1}(h^{t-1},\omega^{t-1})=\hat{a}_{t-1}\}}. \]
 It is immediate to verify that the kernel is well-defined.\medskip 
 
 Two remarks are in order. First, since we consider histories $(h,\omega) \in H\Omega$, we already have that 
 \[ p_1(h_1,\omega_1)\mu_1(\hat{a}_1|\hat{a}^0,h^1,\omega^1) \times \dots \times
p_{t-1}(h_{t-1},\omega_{t-1}|a_{t-1},h^{t-2},\omega^{t-2})p_t(h_t,\omega_t|a_t,h^{t-1},\omega^{t-1})>0 \]
for some $(a_1,\dots,a_t)$. Second, since we only consider feedback rules in the support of $\mu$, all the recommendations with positive probabilities are consistent with the mediation ranges. Hence, players are obedient at these recommendations.\medskip 

We define the conditional probability system on $H\Omega A$ as
\[\beta(h,\omega,\hat{a})= \sum_{f}\beta(f,h,\omega)\indic\{f(h,\omega)=\hat{a}\}.\] 

To complete the proof, we repeat the same steps as in the proof of Theorem \ref{th:equiv}, that is, the additional messages are the recommendations, the kernels $(\xi_1,\dots,\xi_T)$ are the recommendation kernels $(\mu_1,\dots,\mu_T)$, and the conditional probability system is the one on $H\Omega A$ defined above. Since we consider the restriction to recommendations with positive probabilities, the recommendations are consistent with the mediation ranges and players have an incentive to play according to their signals.

\end{document}